\documentclass[superscriptaddress,showpacs,amsmath,amssymb,showkeys,aps,prc,preprint]{revtex4}
\usepackage{hyperref}							  
\usepackage{dcolumn}
\usepackage{bm}
\usepackage{graphicx}
\usepackage[dvips]{color}
\usepackage{epsfig}
\usepackage{float}
\PassOptionsToPackage{caption=false}{subfig} 
\usepackage{subfig}
\usepackage{verbatim}
\usepackage{dcolumn}
\usepackage{setspace}
\usepackage{lineno}
\usepackage{enumitem}
\usepackage{rotating} 
\newcommand{\degree}{\ensuremath{^\circ}}

\begin{document}

\title{Measurement of the elastic scattering cross section of neutrons from argon and neon}

\author{S.~MacMullin}
\email[Corresponding author, E-mail: ]{spm@physics.unc.edu}
\affiliation{Department of Physics and Astronomy, University of North Carolina, Chapel Hill, NC 27599 USA}
\affiliation{Triangle Universities Nuclear Laboratory, Durham, NC 27708 USA}
\author{M.~Kidd}
\altaffiliation[Current address: ]{Tennessee Tech University, Cookeville, TN 38505 USA}
\affiliation{Department of Physics, Duke University, Durham, NC, USA}
\affiliation{Triangle Universities Nuclear Laboratory, Durham, NC 27708 USA}
\author{R.~Henning}
\affiliation{Department of Physics and Astronomy, University of North Carolina, Chapel Hill, NC 27599 USA}
\affiliation{Triangle Universities Nuclear Laboratory, Durham, NC 27708 USA}
\author{W.~Tornow}
\affiliation{Department of Physics, Duke University, Durham, NC, USA}
\affiliation{Triangle Universities Nuclear Laboratory, Durham, NC 27708 USA}
\author{C.R.~Howell}
\affiliation{Department of Physics, Duke University, Durham, NC, USA}
\affiliation{Triangle Universities Nuclear Laboratory, Durham, NC 27708 USA}
\author{M.~Brown}
\altaffiliation[Current address: ]{University of Kentucky, Lexington, KY 40506 USA}
\affiliation{Morehead State University, Morehead, KY, USA}
\affiliation{Triangle Universities Nuclear Laboratory, Durham, NC 27708 USA}



\begin{abstract}
\begin{description}[leftmargin=0in]
\item[Background: ] The most significant source of background in direct dark matter searches are neutrons that scatter elastically from nuclei in the detector's sensitive volume. Experimental data for the elastic scattering cross section of neutrons from argon and neon, which are target materials of interest to the dark matter community, were previously unavailable. 
\item[Purpose: ] Measure the differential cross section for elastic scattering of neutrons from argon and neon in the energy range relevant to backgrounds from ($\alpha,n$) reactions in direct dark matter searches. 
\item[Method: ] Cross-section data were taken at the Triangle Universities Nuclear Laboratory (TUNL) using the neutron time-of-flight technique. These data were fit using the spherical optical model. 
\item[Results: ] The differential cross section for elastic scatting of neutrons from neon at 5.0 and 8.0 MeV and argon at 6.0 MeV was measured. Optical-model parameters for the elastic scattering reactions were determined from the best fit to these data. The total elastic scattering cross section for neon was found to differ by 6\% at 5.0 MeV and 13\% at 8.0 MeV from global optical-model predictions. Compared to a local optical-model for $^{40}$Ar, the elastic scattering cross section was found to differ from the data by 8\% at 6.0 MeV. 
\item[Conclusions: ] These new data are important for improving Monte-Carlo simulations and background estimates for
direct dark matter searches and for benchmarking optical models of neutron elastic scattering from these nuclei.
\end{description}
\end{abstract}

\maketitle

\section{Introduction}

The next generation of low-background physics experiments, including direct detection of Weakly Interacting Massive Particle (WIMP) dark matter may provide significant insight to physics beyond the standard model and the nature of the dominant matter constituent of the universe~\cite{Pri88,Smi90}. The successful detection of WIMPs will provide valuable information about the nature of dark matter. Several current and next generation large-scale detectors designed to search for WIMP dark matter make use of liquified noble gas (Ne, Ar, Xe). Descriptions of several experiments of this type may be found in Refs.\cite{Bou06,Apr11,Ben08,Har10}. These experiments will search for the scintillation light and in some cases the ionization charges generated from the recoiling nucleus after a putative WIMP-nucleus scatter. On the other hand, neutrons having energies less than about 10 MeV may be produced in such experiments via alpha particles produced in decays of $^{238}$U and $^{232}$Th in the detector and surrounding materials that undergo $(\alpha,n)$ reactions. One particular site of these reactions is in borosilicate photomultiplier tube (PMT) glass, because the $^{11}$B$(\alpha,n)^{14}$N cross section is considerably larger than $(\alpha,n)$ cross-section data for other typical detector and shielding materials. PMTs are a common technology for detecting scintillation light in these noble-gas detectors. Neutron-induced nuclear recoils are a particularly dangerous background for these experiments, since they can mimic WIMP events. 

The DEAP/CLEAN experimental program uses large volumes of liquid argon or neon to search for WIMPs from the scintillation channel only~\cite{Bou04,Bou08,Mck07}. Both materials show excellent scintillation properties and make use of pulse-shape discrimination to separate electron and nuclear recoils~\cite{Lip08,Nik08}. The lack of elastic scattering cross-section measurements for neutrons in argon and neon poses a problem for background estimates, performed primarily via Monte Carlo calculations. Precise differential neutron scattering cross sections are required to determine the background contribution due to neutrons in these experiments. Both total and differential cross sections for elastic scattering are important for dark matter experiments, as the rate of multiple scattering inside the detector's sensitive volume is determined by the angular differential cross section. Neutrons that multiple scatter can be cut using event-reconstruction techniques, but calibrations and Monte Carlo simulations using known cross sections are required to determine the efficacy of these analysis cuts. 

We have performed measurements and optical-model fits for the neutron elastic scattering cross section for neon at $E_{n}$ = 5.0 and 8.0 MeV and for argon at $E_n$ = 6.0 MeV for use as inputs to quantitative estimates of neutron backgrounds in these experiments. These neutron energies were determined by where the $^{238}$U and $^{232}$Th induced $\alpha$--neutron yields are the largest in boron \cite{Mei09}. These measurements also provide additional data to benchmark the optical model in a nuclear mass where the model is not well constrained, and will enhance nuclear databases. 

\section{Experimental Technique}

The Neutron Time-of-Flight (NTOF) facility at the Triangle Universities Nuclear Laboratory (TUNL)~\cite{tunl}, located on the campus of Duke University, in Durham, NC, USA was used to measure the neutron elastic scattering cross section of argon and neon. The 10 MV Tandem Van de Graaff accelerator is capable of producing pulsed, mono-energetic neutron beams from about 5-15 MeV using the $^2$H$(d,n)^3$He reaction~\cite{Hut07}. Neutrons were generated by bombarding a 3.15-cm long deuterium gas cell with a deuteron beam pulsed at 2.5 MHz and beam bunches of 2 ns full width at half maximum. Typical deuteron beam current on target for this experiment was about 1.0 $\mu$A. The deuterium gas cell pressure (4.0 atm for $E_n$ = 5.0 and 6.0 MeV, and 7.8 atm for $E_n$ = 8.0 MeV) was chosen to minimize the energy spread in the neutron beam caused by energy loss of the incident deuterons due to ionization in the gas while producing an adequate neutron flux on target to carry out these measurements. The choice of gas pressure allowed a neutron energy spread of about 400 keV. The scattering samples were suspended from thin steel wires located about 10 cm from the neutron production cell and were aligned with the beam line.

The gas target cell is described fully in Ref.\cite{Rup09}. The cell was a 21.0-mm diameter stainless steel sphere with 0.5-mm wall thickness. It was filled with 99.999\% natural neon or natural argon using a coupling device connected to a high-pressure filling station at TUNL. The measurements were carried out with the cell pressurized to 170 atm absolute, which corresponds to a steel-to-gas ratio of about 4-to-1 by mass and is well below the maximum pressure rating of 550 atm. The achievable pressure was limited by the ability to condense the gas in the cell at the filling station. The number of nuclei in the cell was determined by measuring the mass to 0.01 mg accuracy before and after filling. The pressure in the cell was stable, with no measurable difference during the experimental runs, sometimes up to 48 hours. The argon gas cell was filled only once at the beginning of the experiment and the same cell was used throughout. 

\begin{table}[htp]
\centering    
\caption{Description of the gas targets. The diameter is the value of the inner diameter of the sphere.}
\begin{tabular}{l c c c c c} 
\hline
\hline
  Sample   & \multicolumn{3}{c}{Isotopic composition}       & Diameter               &   Number of   \\
           &    \multicolumn{3}{c}{(\% of nuclei)}          &    (cm)              &     sample nuclei   \\
\hline
\hline
  Neon     & $^{20}$Ne  & $^{21}$Ne &$^{22}$Ne                &    2.05   &  1.45--1.94$\times10^{22}$ \\
           & 90.48      & 9.25      & 0.27                             &       &                       \\
\hline
  Argon    &  $^{40}$Ar & $^{36}$Ar & $^{38}$Ar              &   2.05    &  2.40$\times10^{22}$  \\
           & 99.6       & 0.34      & 0.07                             &          &                   \\
\hline
\hline
\end{tabular} 
\label{tab:ntof_gascell}
\end{table}

\begin{figure}
  \centering
  \includegraphics[width=0.75\textwidth]{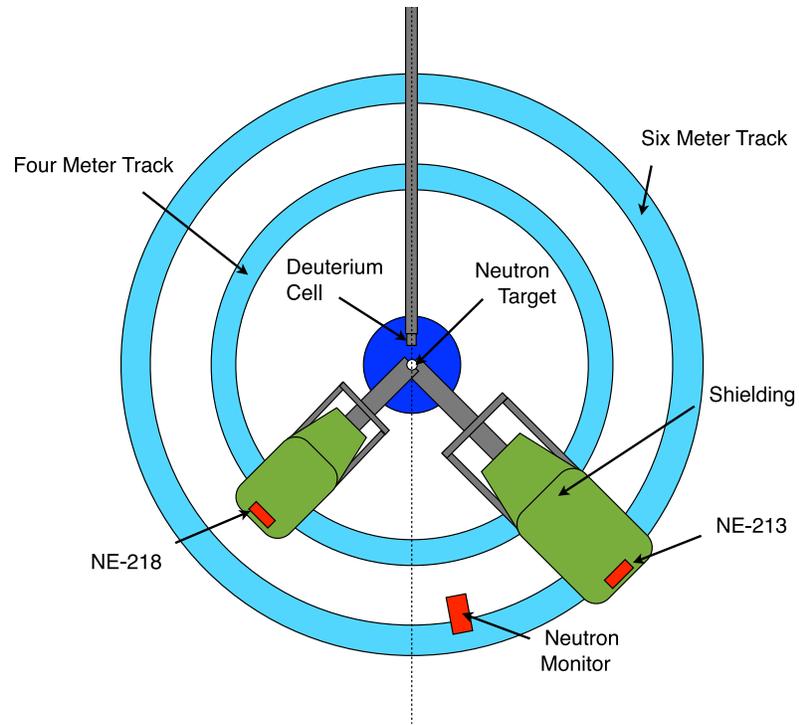}
  \caption{(Color online) The TUNL time-of-flight setup. The target is located 10 cm from the end of the beam line where the neutron production cell is located.  The two neutron detectors move on a four-meter and six-meter track around the neutron target. Figure not to scale.}
\label{fig:tofsetup}
\end{figure}

The scattered neutrons were detected by two liquid scintillator detectors, placed on both sides of the beam axis. The detectors were collimated and shielded from external backgrounds by copper, lead and polyethylene. Each of the two neutron detectors was located at the end of the collimator inside the shielding which was mounted to a steel carriage that moved around the central scattering target on tracks. The detectors were positioned about four meters and six meters from the central axis. The detector on the four-meter track was filled with NE-218 liquid scintillator, and the detector on the six-meter track was filled with NE-213 liquid scintillator (Nuclear Enterprise Ltd., Edinburgh, UK). Each liquid scintillator was optically coupled to a PMT. A third liquid scintillator detector was placed behind the 6-meter track at about 10$^{\circ}$ with respect to the beam axis. This detector was unshielded and had a direct line-of-sight to the neutron production cell behind the scattering target. This detector was used to monitor the direct neutron and $\gamma$-ray spectra and normalize the yields from the four- and six-meter detectors. A schematic of the TUNL time-of-flight setup is shown in Fig.~\ref{fig:tofsetup} and technical details on the detectors are shown in Table~\ref{tab:NE21X}. 

\begin{figure}   
	\centering
	\subfloat{\label{fig:PHPSD}\includegraphics[width=0.65\textwidth]{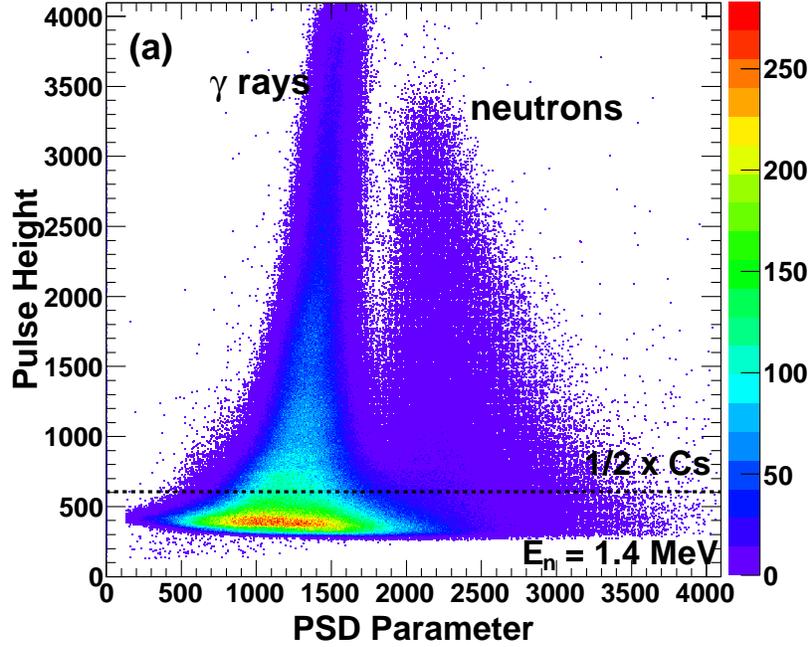}}\\
	\subfloat{\label{fig:PH}\includegraphics[width=0.65\textwidth]{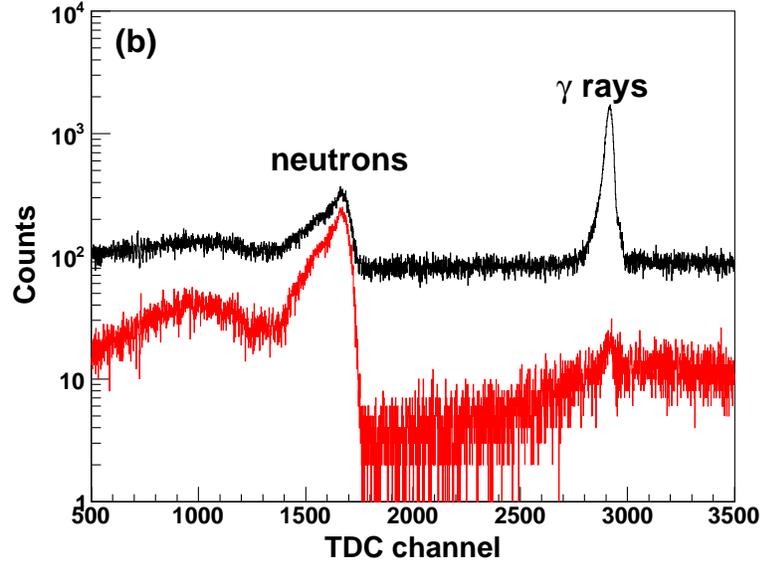}}
	\caption{(Color online) An example PSD and TDC spectrum from the six-meter detector for 6.0-MeV neutrons scattered from argon at $\theta_{CM}$ = 111.4$^{\circ}$. (a) A PSD spectrum showing the pulse-height output from the PMT versus the fast/slow component of the scintillator light output. This allows $\gamma$ ray and neutron interactions in the scintillator to be separated. An energy threshold cut for $1/2\times$Cs ($E_n$ = 1.4 MeV) is also shown. (b) A TDC spectrum with no cuts (black) and with both the PSD and threshold cut (red) are also compared. Gamma rays and low-energy neutrons are removed with the PSD and threshold cuts leaving only neutrons from elastic and inelastic scattering in target cell as well as ``room-return'' neutrons. Time of flight increases with decreasing channel number. The TDC has a gain of 0.088 ns/channel.}
\label{fig:psd}
\end{figure}

A beam pick-off signal was generated using a capacitive pick-off (CPO) unit located in the beam line just before the deuterium gas cell. As a beam bunch passed through the capacitor, a charge pulse was used to define the ``start signal'' for a neutron time-of-flight measurement. The signals from each of the neutron detectors were processed identically. The PMT anode signals were sent to a Mesytec MPD-4 n-$\gamma$ discriminator~\cite{mpd4}. Because liquid scintillator detectors are sensitive to both neutrons and $\gamma$ rays, it was necessary to discriminate against $\gamma$ rays to reduce backgrounds. This was accomplished using pulse-shape discrimination (PSD) techniques. Gamma rays interact directly with atomic electrons in the scintillator and neutrons interact with protons via $n-p$ scattering. Because the photon emission decay rate in the scintillator is shorter for electrons, which are low-ionizing, compared to recoiling protons, which are high-ionizing, the decay time of the resulting PMT signal is shorter for $\gamma$-ray interactions than it is for neutron interactions. The MPD-4 module output a logic pulse at a constant fraction of the trailing edge of the PMT anode signal, thus outputting a smaller pulse-height for a $\gamma$ ray than for a neutron signal. Additional cuts were made at an energy threshold corresponding to $1\times$Cs ($E_n$ = 2.2 MeV) for the 8.0-MeV data and at $1/2\times$Cs ($E_n$ = 1.4 MeV) for the 5.0- and 6.0-MeV data, where 1$\times$Cs corresponds to the Compton edge (447 keVee) from a backscattered 662-keV $\gamma$ ray from a $^{137}$Cs source. An example PSD spectrum is shown in Fig.~\ref{fig:psd}.

\begin{figure}   
	\centering
	\subfloat{\label{fig:inout}\includegraphics[width=0.65\textwidth]{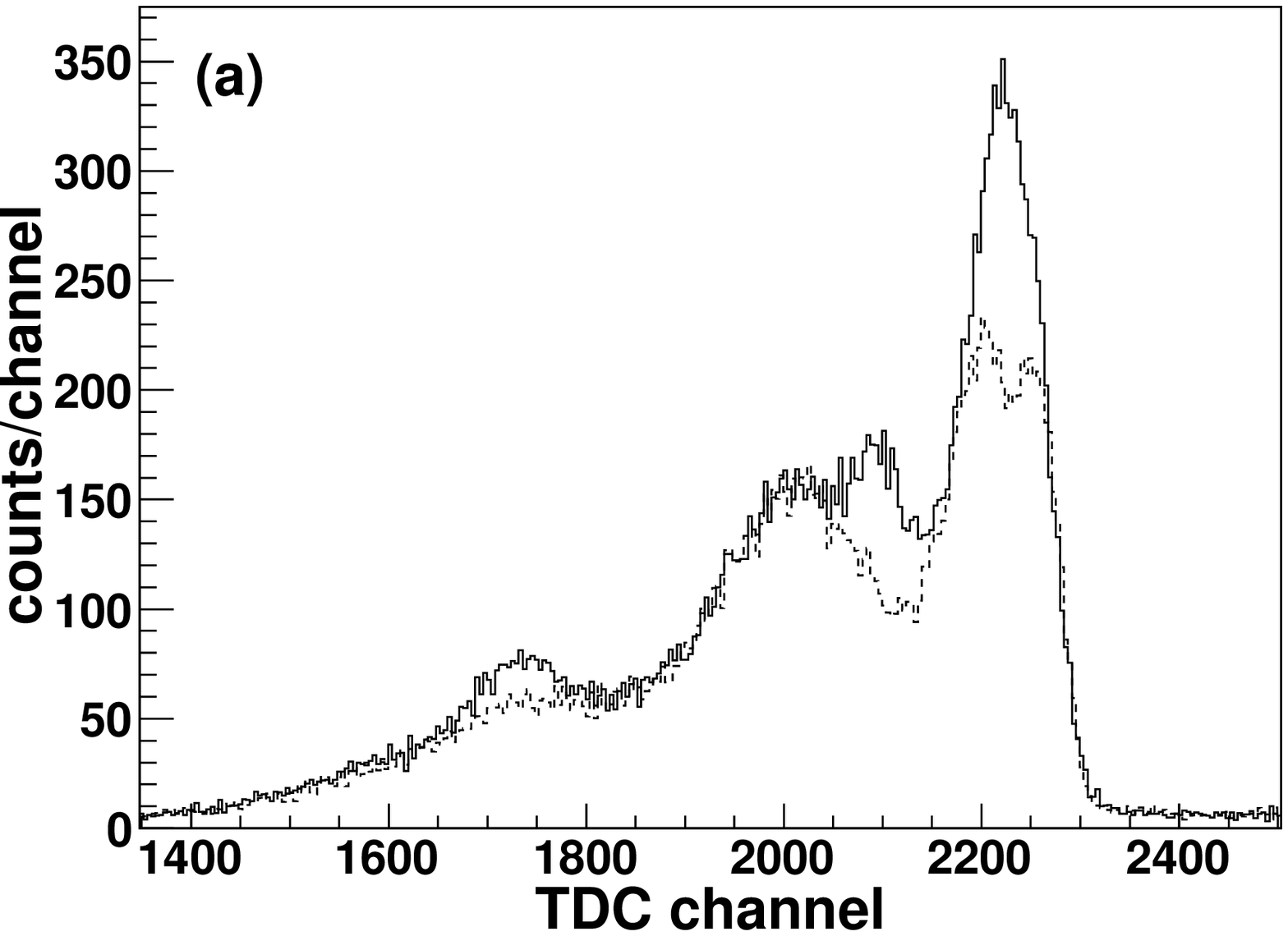}}\\
	\subfloat{\label{fig:diff}\includegraphics[width=0.65\textwidth]{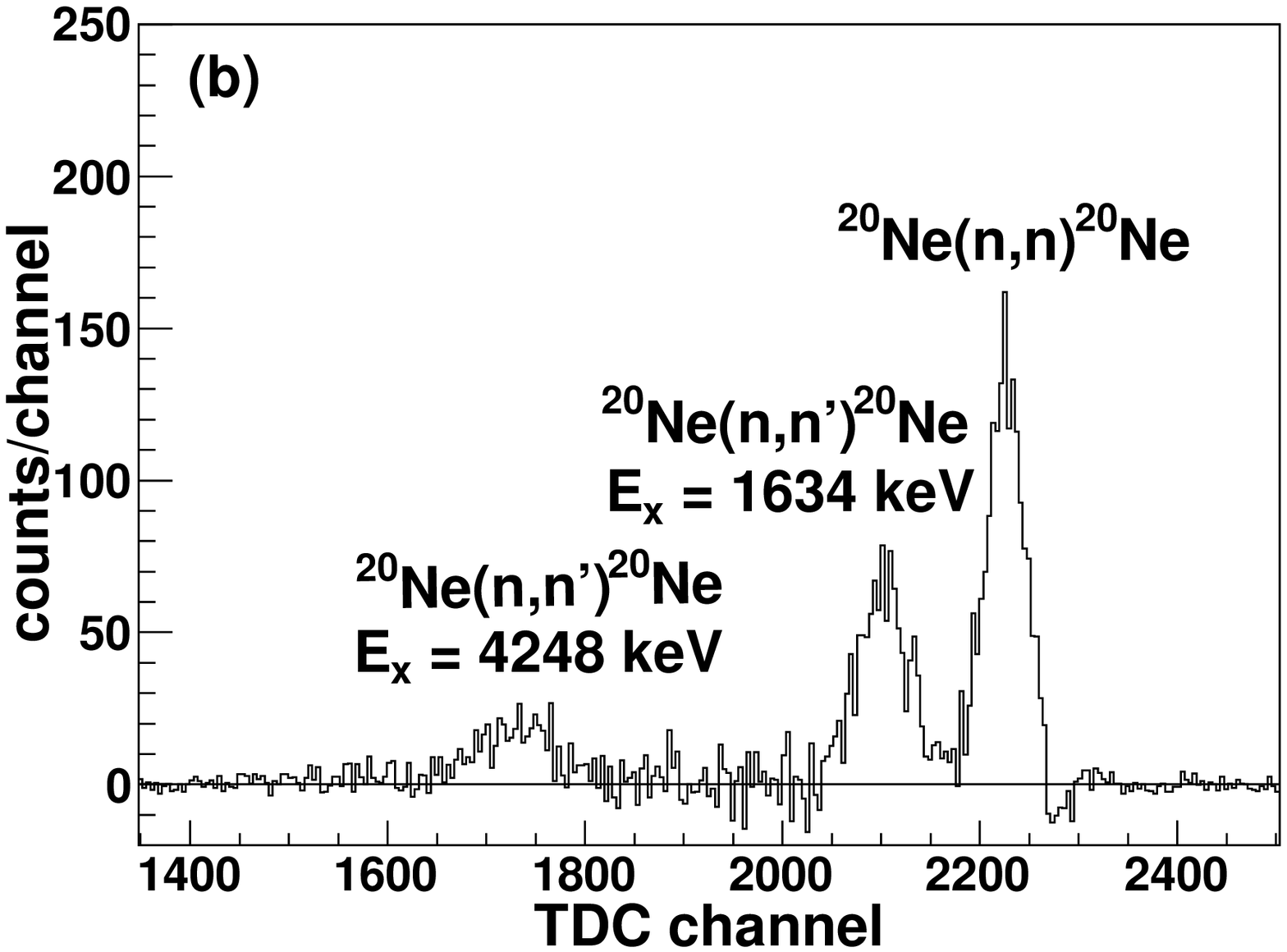}}
	\caption{Time-of-flight spectrum from the four-meter detector for 8.0-MeV neutrons scattered from neon at $\theta_{CM}$ = 102.9$^{\circ}$. The flight path length was 317.5 cm. Time of flight increases with decreasing channel number. The TDC has a gain of 0.088 ns/channel. (a) The ``sample in" (solid) and ``sample out" (dashed) normalized to the neutron monitor. (b) The difference where the $0^+$ (elastic), $2^+$ (1634-keV) and $4^+$ (4248-keV) states in $^{20}$Ne are visible.}
\label{fig:tof_spectrum}
\end{figure}

A run was done after each ``sample in" measurement with the unfilled gas cell so that backgrounds could be subtracted. The background subtracted yields were corrected for dead time and normalized to the neutron counter that had a direct line of sight to the neutron production cell. Data were also normalized to the integrated beam current to check for consistency. A typical time-of-flight spectrum is shown in Fig.~\ref{fig:tof_spectrum}. The elastic and two inelastic scattering peaks in neon are clearly visible. To determine the cross section, the data were normalized to $n-p$ scattering using a small cylindrical polyethylene (C$_2$H$_4$)$_n$ target. Backgrounds from elastic and inelastic scattering from carbon were subtracted using a graphite target which was made to have the same number of carbon atoms as the polyethylene.  

\clearpage

\section{Cross-Section Analysis and Results}

For each angle at a given neutron energy, a differential elastic scattering cross section was obtained using

\begin{equation}
    \frac{d\sigma}{d\Omega}(E_n,\theta_s) = \frac{Y_s(\theta_s)}{Y_p(\theta_p)} \frac{A_p(\theta_p)}{A_s(\theta_s)} \frac{\epsilon_p}{\epsilon_s} \frac{n_p}{n_s} \frac{d\sigma_p}{d\Omega}(E_n,\theta_p),
    \label{eqn:cs}
\end{equation}

\noindent where $\theta_s$ and $\theta_p$ are the angles, relative to the beam axis, of the scattering sample and polyethylene normalization sample, respectively. $Y_s(\theta_s)$ and $Y_p(\theta_p)$ are the background subtracted and dead-time corrected time-of-flight yields of the scattering sample and hydrogen in the polyethylene normalized to the neutron flux monitor detector. $A_s(\theta_s)$ and $A_p(\theta_p)$ account for attenuation and multiple scattering in the scattering sample and polyethylene. The quantities $\epsilon_s$  and $\epsilon_p$ are the detection efficiencies of a neutron scattered elastically from a target nucleus and a hydrogen atom in the polyethylene. The quantities $n_s$  and $n_p$ are the number of target nuclei in the scattering sample and the number of hydrogen atoms in the polyethylene. The $n-p$ normalization cross section was obtained from the Nijmegen partial-wave analysis of $N-N$ scattering data~\cite{Sto93,nnonline}.

\subsection{Detection Efficiency}

\begin{centering}
\begin{figure}[htp]
  \centering
  \subfloat{\label{fig:4meff}\includegraphics[width=0.65\textwidth]{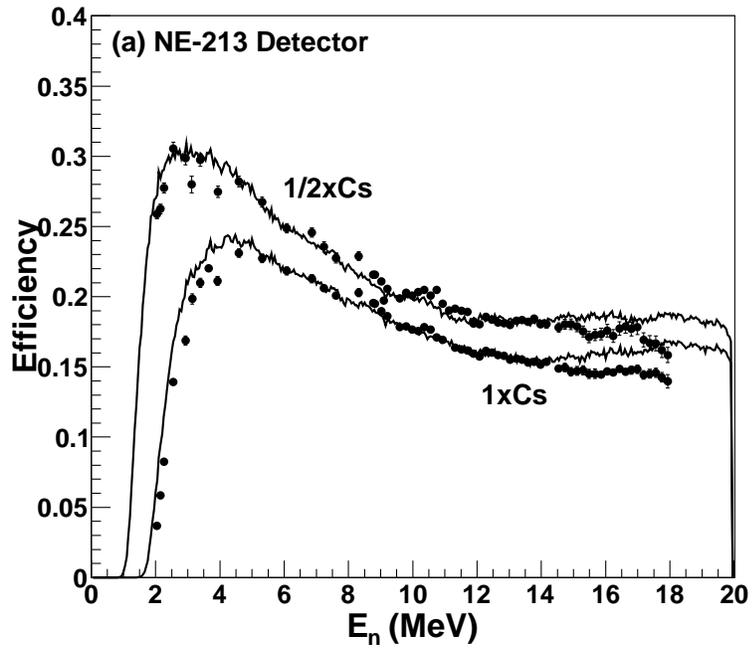}}\\             
  \subfloat{\label{fig:6meff}\includegraphics[width=0.65\textwidth]{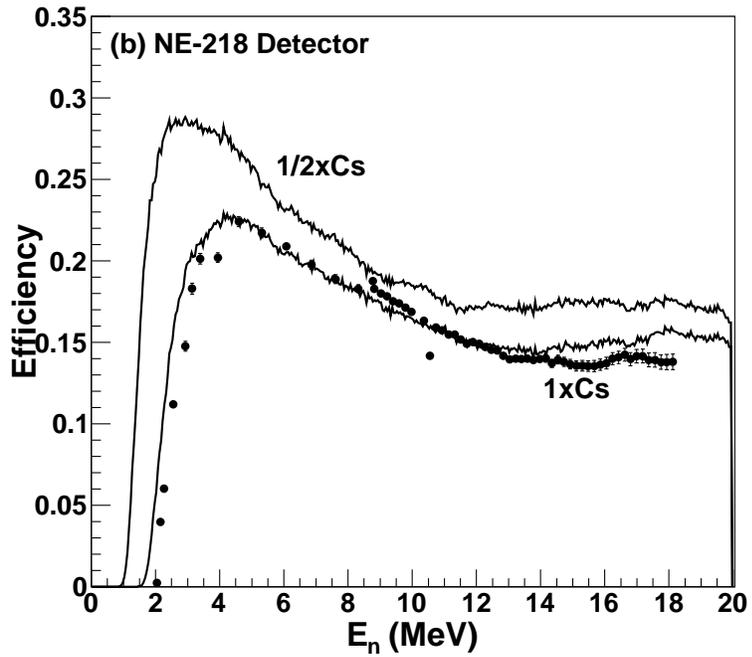}}
  \caption{Efficiency curves for the TUNL time-of-flight liquid scintillator detectors at thresholds of $1\times$Cs and $1/2\times$Cs. The solid circles are data from Pedroni~\cite{Ped86}. The solid curves are the results from the \textsc{neff}7~\cite{Die82} simulation.}
  \label{fig:eff}
\end{figure}
\end{centering}

The neutron response and detector efficiencies were simulated using the code \textsc{neff}7~\cite{Die82}. Simulations for both detectors were performed using various neutron energy thresholds. The simulation code was written for NE-213 liquid scintillators and was modified for the NE-218 detector by changing the scintillator density and hydrogen-to-carbon ratio, see Table~\ref{tab:NE21X}. The simulations were performed for neutron energy thresholds corresponding to 1$\times$Cs and 1/2$\times$Cs. The results of the simulation are shown in Fig.~\ref{fig:eff} and are compared to available data. It should be noted that because the efficiency enters the cross-section formula as a ratio between the detection efficiency of a neutron scattering from a target nucleus and from hydrogen in the polyethylene, the analysis is not affected by the absolute efficiency but only by the shape of the efficiency curve.

\begin{sidewaystable}[htp]
\centering    
\caption{Properties of the TUNL time-of-flight liquid scintillator detectors. Both detectors are right cylinders. Densities and hydrogen-to-carbon ratios are taken from Ref.~\cite{Ang79}.}
\begin{tabular}{l c c c c c c} 
\hline
\hline
        &   Radius    & Thickness  &   Density     &  $N_H$            & $N_C$             & hydrogen-to-carbon  \\
        &   (cm)       &    (cm)    &   (g/cm$^3$)  & (hydrogen atoms/cm$^3$)  & (carbon atoms /cm$^3$) & ratio\\
\hline
NE-213 (six-meter track) &   6.35    &   5.08        &   0.874       & 4.82 $\times 10^{22}$ & 3.98 $\times 10^{22}$ & 1.213 \\
NE-218 (four-meter track) &   4.45    &   5.08        &   0.879       & 5.10 $\times 10^{22}$ & 3.99 $\times 10^{22}$ & 1.28 \\       
\hline
\hline
\end{tabular}
\label{tab:NE21X}
\end{sidewaystable}

\subsection{Finite-Geometry and Multiple-Scattering Corrections}

The quantity $A_p$ was determined using the ``disc approximation'' from Kinney~\cite{Kin70} and is given by

\begin{gather}
    \label{eqn:Ap1}
    A_p(\theta_p) = \exp(\frac{\pi}{4} \Sigma(E_0)R + \frac{8}{3\pi} \Sigma(E_1)R) \\
    \notag \Sigma = n_H\sigma_H + n_C\sigma_C,
\end{gather}  

\noindent where $E_0$ is the incident neutron energy and $E_1$ is the neutron energy after a single scatter at $\theta_p$. The radius of the polyethylene target is $R$, and $n_{H,C}$ denote the number of hydrogen and carbon atoms in the polyethylene target, respectively. The total neutron cross section for hydrogen  is $\sigma_H$ and $\sigma_C$ is the non-elastic neutron cross section for carbon. The quantity $\sigma_H$ were calculated using Ref.~\cite{nnonline} as described above and $\sigma_C$ was taken from the ENDF/B-VII.0 database~\cite{Cha06}. 

The scattering data were corrected for multiple scattering, attenuation and finite geometry effects in the target gas and steel cell with a custom C++-based Monte Carlo simulation in which single simulated neutrons were tracked as they traversed the simulated target and cell geometry. The simulation included all the relevant materials' elastic differential cross-section data, densities and dimensions. The simulation also included a neutron beam profile and angular distribution based the incident deuteron beam profile on the production cell convoluted with the $^{2}$H($d,n$)$^{3}$He cross section. It returned a ``measured" cross section for given input cross section for the target. By taking the appropriate ratio between the measured and input cross sections, we computed a correction factor. This correction was only performed once using the measured data as input, but for some test cases the corrected cross section was used an the input for a second iteration. These second iterations showed a significantly smaller correction, as expected, and provided an estimate for the residual uncertainty in the multiple scattering and geometry corrections. The effect on the correction factor $A_{s}$ due to variation in the number of sample nuclei was negligible. The corrections were less than about 10\% at most angles but were as high as 25\% at the forward angle minimum of $\sigma(\theta)$ for neon at 8.0 MeV.

\subsection{Measurement Uncertainty}

An uncertainty of 3\% was assigned to the detector efficiency based on the agreement of the data with the~\textsc{neff7} simulation in the neutron energy range used in this experiment. Similar agreement was found using the same simulation code in Ref.~\cite{Tro09}. The number of target nuclei in the gas cell was measured by weighing the filled and unfilled gas cell on a precision balance. The systematic uncertainty in the number of target nuclei in the scattering sample was 0.2\%. The uncertainty in the number of hydrogen atoms in the polyethylene was 0.7\%. An uncertainty of 3\% in the polyethylene correction factor $A_p$ was determined from the uncertainties in the target radius and hydrogen and carbon densities in the polyethylene. The contribution of the uncertainties in the hydrogen and carbon cross sections in Eqn.~\ref{eqn:Ap1} to the total uncertainty in $A_p$ was negligible. An uncertainty of 1\% was assigned to $A_s$ based on the statistical fluctuation in the correction factor. An uncertainty of 0.5\% was assigned to the $n-p$ normalization cross section based on agreement between several different models and analyses of $N-N$ scattering data available from Nijmegen~\cite{nnonline}.

The statistical uncertainty played the largest role in this experiment. Because there were about four times as many iron nuclei in the gas cell as target nuclei, a large background was subtracted for each angle. We attempted to collect enough scattering events at each angle so that the statistical uncertainty was around 10\%. At points where the cross section was small, or where the elastic scattering cross section for $^{56}$Fe (from the stainless steel gas cell) was much larger than the argon or neon cross section, the statistical uncertainty was as high as 20\%. The systematic and statistical uncertainties are summarized in Table~\ref{tab:uncertainties}.

\begin{table}[htp]
    \centering
\caption{Systematic and statistical uncertainties.}
\begin{tabular}{l c} 
\hline
\hline
\multicolumn{2}{c}{Systematic Uncertainties} \\
\hline
multiple scattering correction factor, $A_s$  & 1\% \\ 
polyethylene correction factor, $A_p$   & 3\%\\
detector efficiency, $\epsilon_s$/$\epsilon_p$            & 3\% \\
number of target nuclei, $n_p$/$n_s$        & 0.7\% \\
$n-p$ cross section         & 0.5\%\\
\hline
Total systematic uncertainty    & 4.4\% \\
\hline
\hline
\multicolumn{2}{c}{Statistical Uncertainties} \\
\hline
$Y_s$   &   5--20\%\\
$Y_p$   &   1--4\%\\
\hline
\hline
\end{tabular}
\label{tab:uncertainties}
\end{table}

\subsection{Experimental Results}

Experimental results are given in Figs.~\ref{fig:cs_argon}~and~\ref{fig:cs_neon} and Tables~\ref{tab:ntof_neon5_data}--\ref{tab:ntof_argon14_data}. These data were fit with a Legendre polynomial expansion given by

\begin{equation}
    \frac{d\sigma}{d\Omega}(E_n,\theta_s) = \sum_{l=0}A_l(E_n)P_l(\cos\theta_s),
\end{equation}

\noindent where $A_l(E_n)$ were free parameters. The maximum value for $l$ was determined from the condition that the $\chi^2$/NDF of the fit for the next order ($l+1$) was greater than for the current fit ($l$). 

It is useful to have a point at zero degrees where the differential cross section is largest. Wick's limit~\cite{Wic43,Wic49} sets a lower limit on the differential elastic scattering cross section at zero degrees by relating the zero-degree cross section to the total neutron cross section using the optical theorem. Although this quantity only represents the contribution from the imaginary part of the scattering amplitude, it is known to be nearly an equality~\cite{Die03,Coo58}. These values, which are listed in Tables~\ref{tab:ntof_neon5_data}--\ref{tab:ntof_argon14_data}, were based on an extrapolation of the total neutron scattering data of Vaughn {\it et al.}~\cite{Vau60} and were included as a data point in each fit. 

The elastic scattering cross section for neon, determined from the Legendre polynomial fits, was found to be 1290 $\pm$ 40 mb for $E_n$ = 5.0 MeV and 940 $\pm$ 30 mb for $E_n$ = 8.0 MeV. There was one angular distribution for elastic scattering of neutrons from $^{40}$Ar at 14.0 MeV available in the literature measured by Beach {\it et al.}~\cite{Bea67}. It is included along with the 6.0-MeV data to help form a complete understanding of the current availability of data and to aid in optical-model predictions. The 14-MeV data was fit using the same procedure as for the time-of-flight data. The elastic scattering cross section for argon was found to be 2170 $\pm$ 60 mb for $E_n$ = 6.0 MeV and 970 $\pm$ 20 mb for $E_n$ = 14.0 MeV.

\section{Discussion}

\subsection{Spherical Optical-Model Analysis}

The experimental data were fit using a spherical optical model (SOM)~\cite{Hod63} using the \textsc{genoa} code~\cite{Per67}. The code performs a searching procedure with up to 10 free parameters to define the potential and uses a numerical procedure to solve the time-independent Schr\"{o}dinger equation with numerical fitting based on the generalized least squares method~\cite{Gui00}.  We used a potential of the same form as in Koning and Delaroche~\cite{Kon03}, which included a Woods-Saxon real volume term, an imaginary surface derivative term and spin-orbit terms:
 
\begin{multline}
    U_{opt}(E_n,r) = 
    -V_0(E_n) f(r,r_0,a_0) + \\
    4ia_iW_d(E_n)\frac{d}{dr}f(r,r_i,a_i) + 
    V_{so}(E_n)\left(\frac{\hbar}{m_{\pi}c}\right)^2\frac{\vec{\sigma} \cdot \vec{l}}{r}\frac{d}{dr}f(r,r_{so},a_{so}) + \\             iW_{so}(E_n)\left(\frac{\hbar}{m_{\pi}c}\right)^2\frac{\vec{\sigma} \cdot \vec{l}}{r}\frac{d}{dr}f(r,r_{so},a_{so})
    \label{eqn:opt}
\end{multline}

\begin{equation*}
    f(r,r_x,a_x) =[1+\exp(r-r_xA^{1/3})/a_x]^{-1}
\end{equation*}    

Because the optical model only describes the direct reaction, in general, the compound nucleus cross section must be subtracted from the data before optical-model fits are attempted. The compound nucleus cross section was calculated using the nuclear reaction code \textsc{talys}~\cite{Kon05}, which used a Hauser-Feshbach statistical calculation with a Moldauer width fluctuation correction factor~\cite{Hod86,Mol76,Mol80}. In light nuclei, analytic expressions for the density of states are not reliable because the level densities are too low and these calculations can only be used to provide an upper limit to the compound nucleus reaction. For neon, below about 10 MeV, the compound nucleus contribution to the cross section was calculated to be large compared to the direct reaction cross section. For this reason, we did not attempt to correct the neon data for the compound nucleus before performing optical-model fits. The compound nucleus corrections for argon at 6.0 MeV were made before describing the data with the optical model. At 14.0 MeV, the compound nucleus cross section is negligible.

The optical-model parameters from the search are shown in Table~\ref{tab:opt}. Using the measured cross section for argon at 6.0 and 14.0 MeV, a parameter set was determined based on the local optical-model parameters from Koning and Delaroche~\cite{Kon03} for $^{40}$Ar. In the SOM, only the potential well depths $V_0$, $W_d$, $V_{so}$ and $W_{so}$ depend on energy. Because of the weak energy dependence of the spin-orbit parameters, only small modifications to the real volume potential ($V_v$) and imaginary surface potential ($W_d$) were needed to describe the data at 6.0 and 14.0 MeV. The surface diffuseness parameters ($a_x$) and radii ($r_x$) depend on the mass numbers only, and were not treated as free parameters in the fit. Because of the good agreement between the optical-model calculations and experimental data, we conclude that the available data were consistent with each other and also consistent with the trends of the $^{40}$Ar local optical-model potential.

Because the Koning--Delaroche global optical model is not expected to perform well for light nuclei such as neon, initial estimates were also taken from Dave and Gould~\cite{Dav83}, which showed good agreement with data for $1-p$ shell nuclei for incident neutron energies from 7 to 15 MeV. In our case, adequate descriptions of the data were only found by allowing the surface diffuseness parameters and radii in the volume and surface terms to vary with energy. These were treated as free parameters in the search. The spin-orbit parameters were taken from Ref.~\cite{Dav83} and were treated as fixed parameters. $W_{so}$ was taken from the Koning--Delaroche global potential since the Dave--Gould potential did not include this term. The differential cross section was calculated for energies not measured using this global fit to the data assuming a linear energy dependence for each parameter. For neon, although the model adequately describes the measured data, we do not expect our global fit to apply outside the $A=20$ mass range and may be limited to the energy range measured in the current experiment.

\begin{table*}[htp]
\centering    
\caption{Spherical optical-model parameters for neutrons incident on argon and neon. Potential well depths are in MeV and radii and surface diffuseness parameters are in fm.}
\begin{tabular}{l c c c c c c c c c c} 
\hline
\hline
	$E_n$ (MeV)		& $V_0$	& $r_0$ & $a_0$ & $W_d$ & $r_i$ & $a_i$ & $V_{so}$ & $W_{so}$ & $r_{so}$ & $a_{so}$		\\
\hline
Argon \\
\hline
6.0                 &   51.814  &   1.188   &   0.670   &   4.579   &   1.290   &   0.543   &   5.77    &   -0.02   &   1.010   &   0.580\\ 
14.0                &   49.827  &   1.188   &   0.670   &   5.180   &   1.290   &   0.543   &   5.59    &   -0.06   &   1.010   &   0.580\\ 
\hline
Neon \\
\hline
5.0                 & 52.340   & 1.242 & 0.518 & 7.840   & 2.014 & 0.179 & 5.550 & -0.030   & 1.150    & 0.500        \\
8.0                 & 45.057   & 1.126 & 0.306 & 7.759   & 1.954 & 0.145 & 5.550 & -0.040   & 1.150    & 0.500        \\
\hline
\hline
\end{tabular}
\label{tab:opt}
\end{table*} 

\begin{centering}
\begin{figure}[htp]
  \centering
  \subfloat{\label{fig:cs_argon_6}\includegraphics[width=0.65\textwidth]{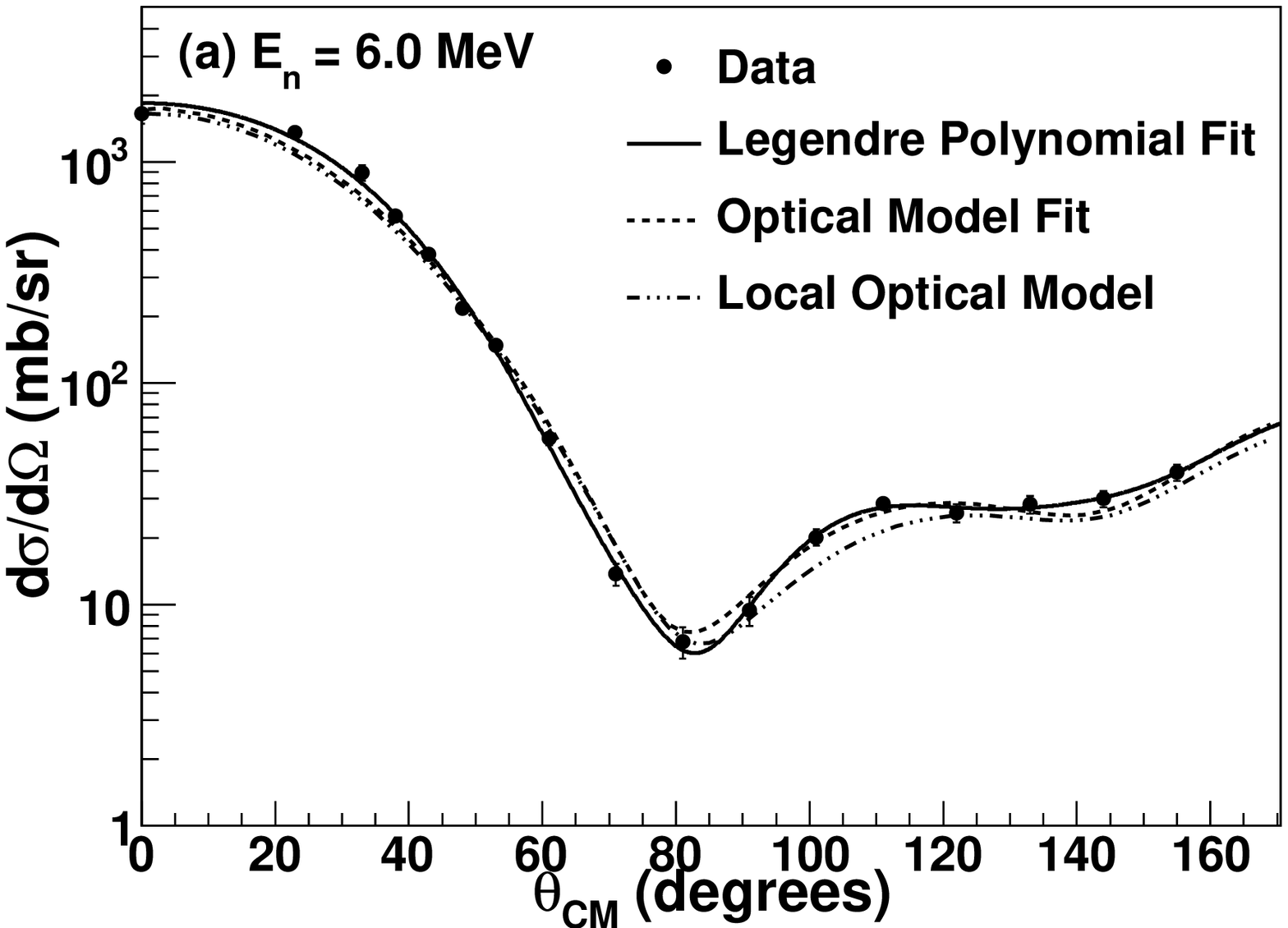}}\\
  \subfloat{\label{fig:cs_argon_14}\includegraphics[width=0.65\textwidth]{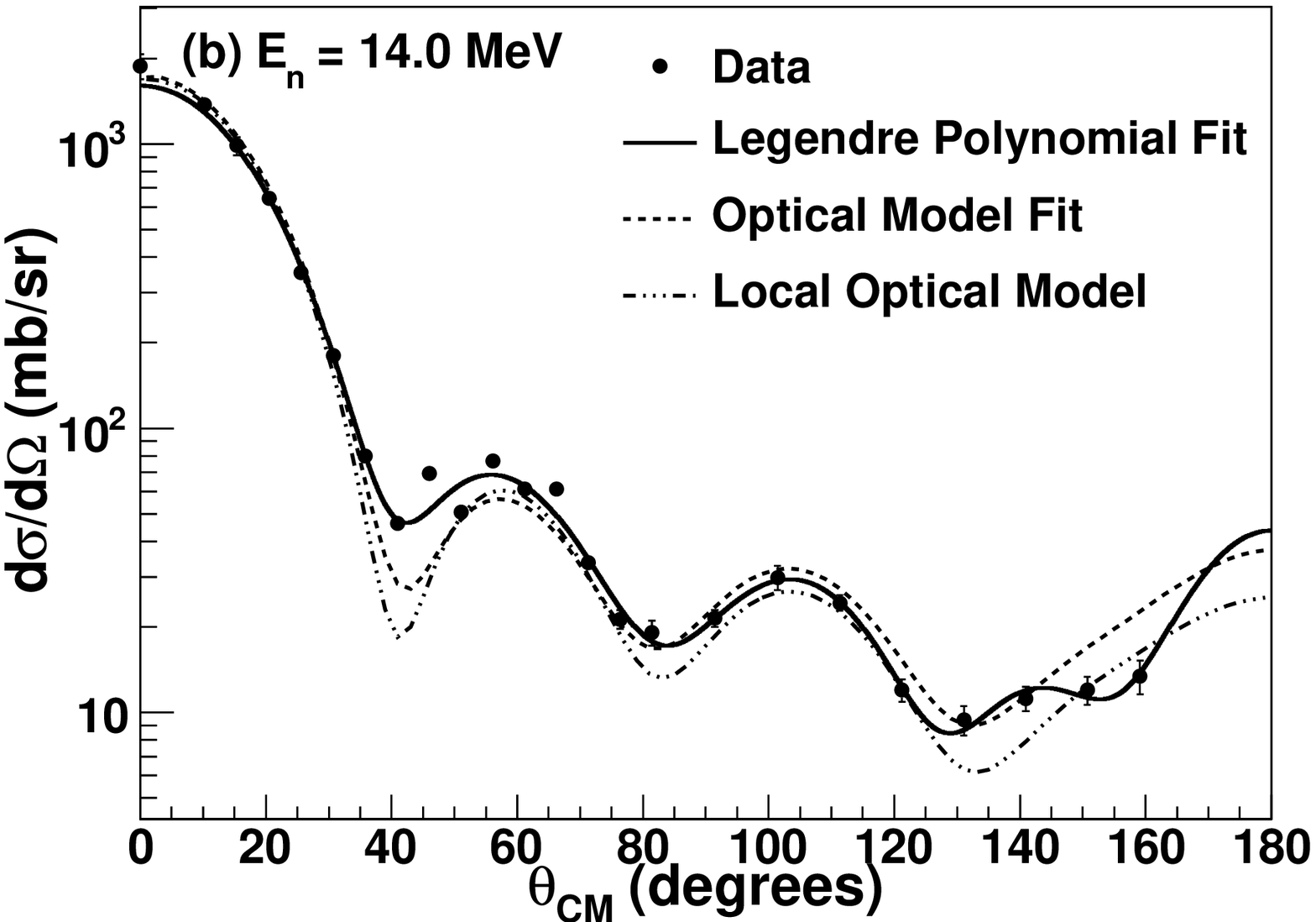}}
  \caption{The differential elastic scattering cross section of neutrons from $^{\rm nat}$Ar. The 6.0-MeV data were fit with a ninth-order Legendre polynomial expansion ($\chi^2$/NDF = 1.6). The 14.0-MeV data~\cite{Bea67} were fit with a tenth-order Legendre polynomial expansion ($\chi^2$/NDF = 6.3). The SOM prediction using the parameters in Table~\ref{tab:opt} are also shown. The data and fits are compared to the local optical-model predictions for $^{40}$Ar($n,n$)$^{40}$Ar using the Koning--Delaroche local potential~\cite{Kon03}.}               
  \label{fig:cs_argon}
\end{figure}
\end{centering}

\begin{centering}
\begin{figure}[htp]
  \centering
  \subfloat{\label{fig:cs_neon_5}\includegraphics[width=0.65\textwidth]{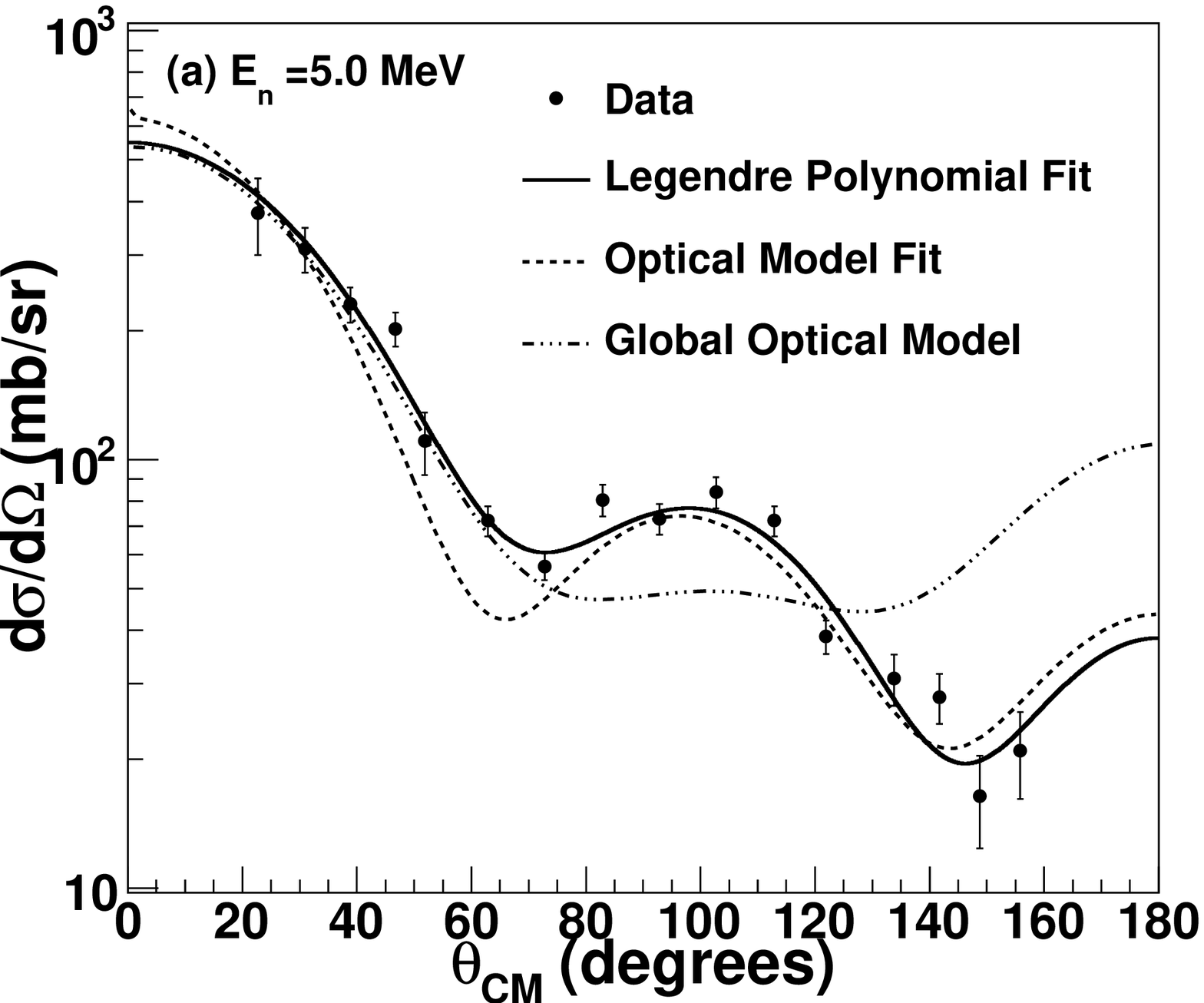}}\\
  \subfloat{\label{fig:cs_neon_8}\includegraphics[width=0.65\textwidth]{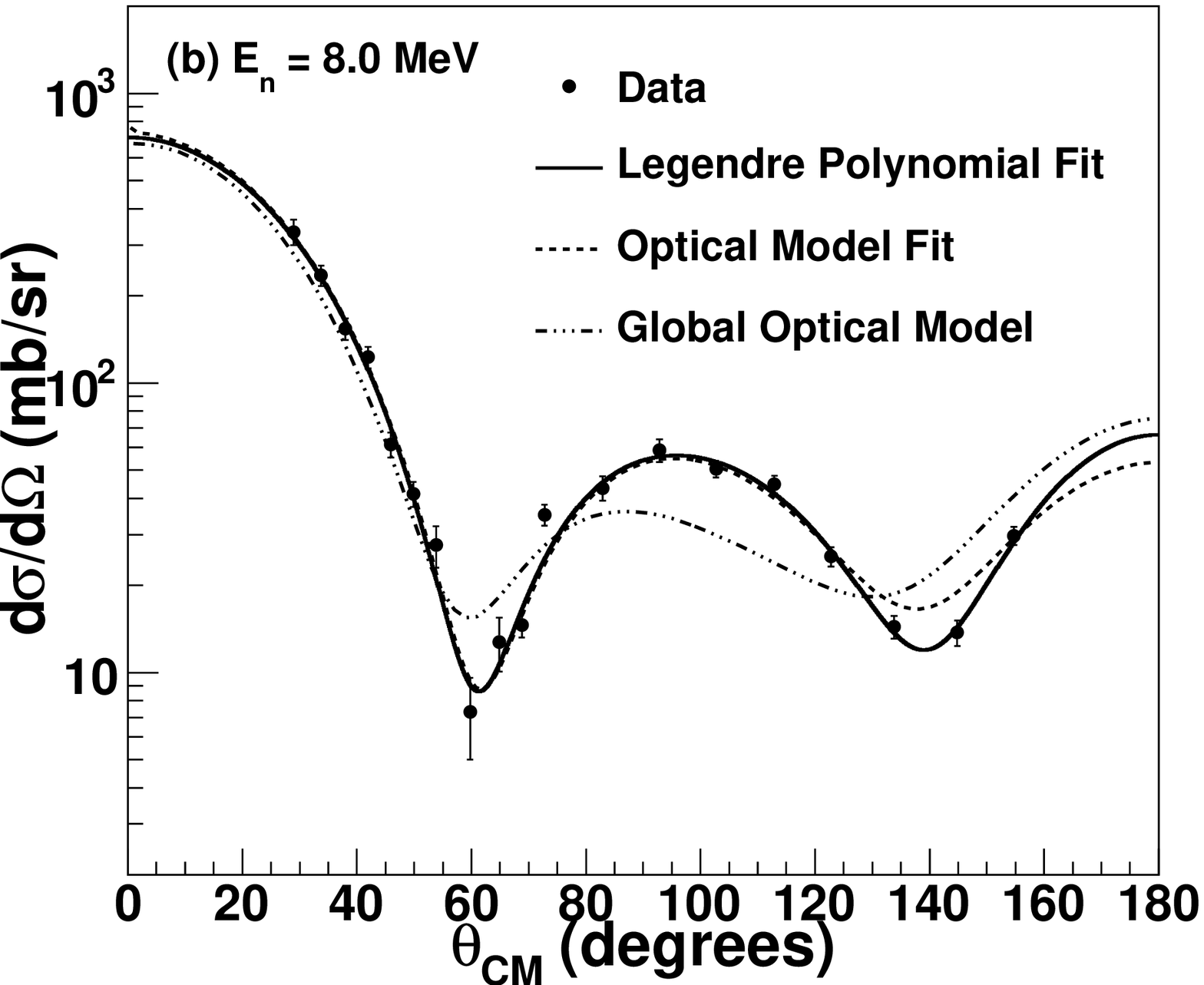}}
  \caption{The differential elastic scattering cross section of neutrons from $^{\rm nat}$Ne. The 5.0-MeV data were fit with a fourth-order Legendre polynomial expansion ($\chi^2$/NDF = 2.2). The 8.0-MeV data were fit with a sixth-order Legendre polynomial expansion ($\chi^2$/NDF = 2.0). The SOM prediction using the parameters in Table~\ref{tab:opt} are also shown. The data and fits are compared to the global optical-model predictions for $^{20}$Ne($n,n$)$^{20}$Ne using the Koning--Delaroche potential~\cite{Kon03}.}               
  \label{fig:cs_neon}
\end{figure}
\end{centering}

\begin{centering}
\begin{figure}[htp]
  \centering
  \subfloat{\label{fig:argon_2d}\includegraphics[width=0.65\textwidth]{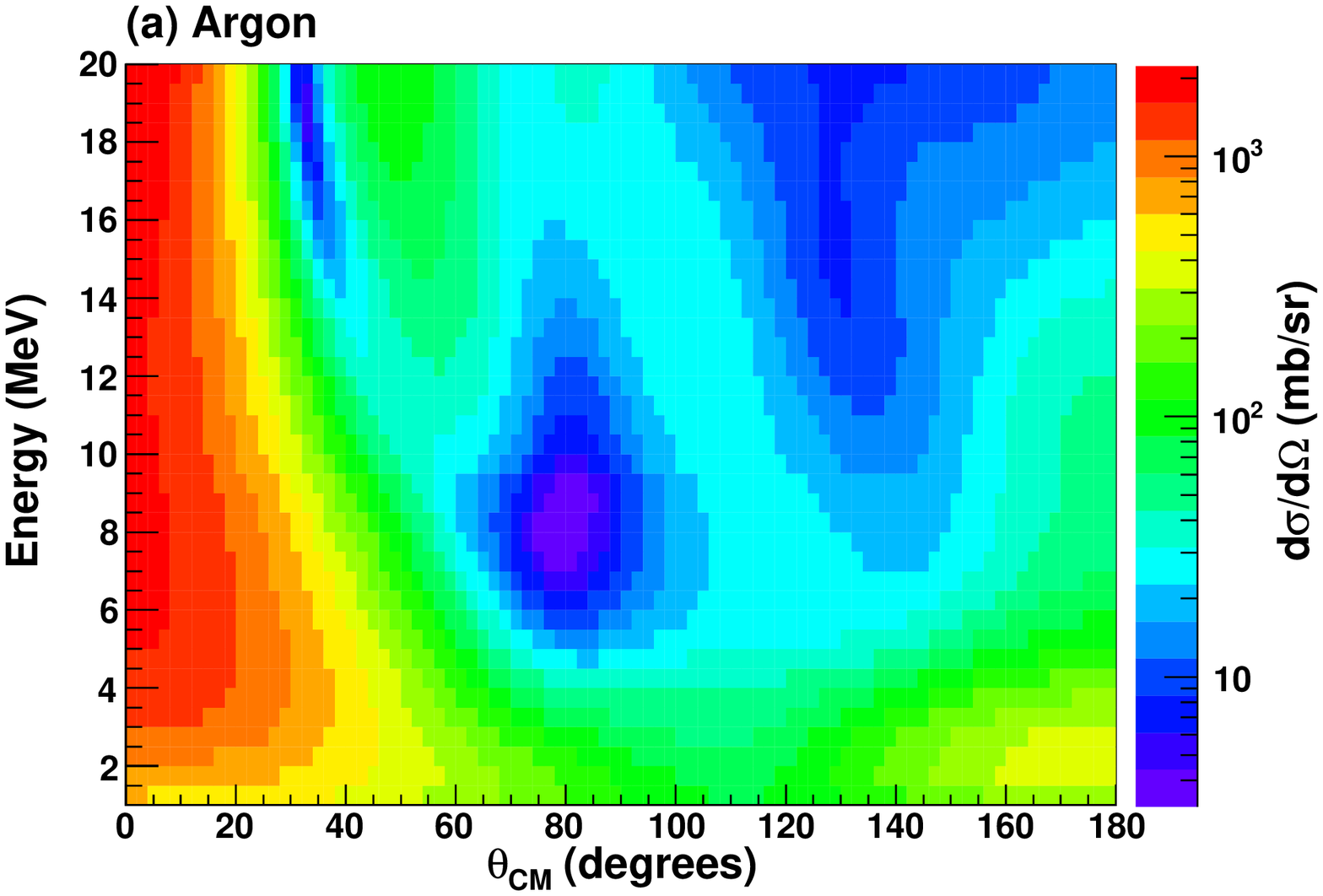}}\\
  \subfloat{\label{fig:neon_2d}\includegraphics[width=0.65\textwidth]{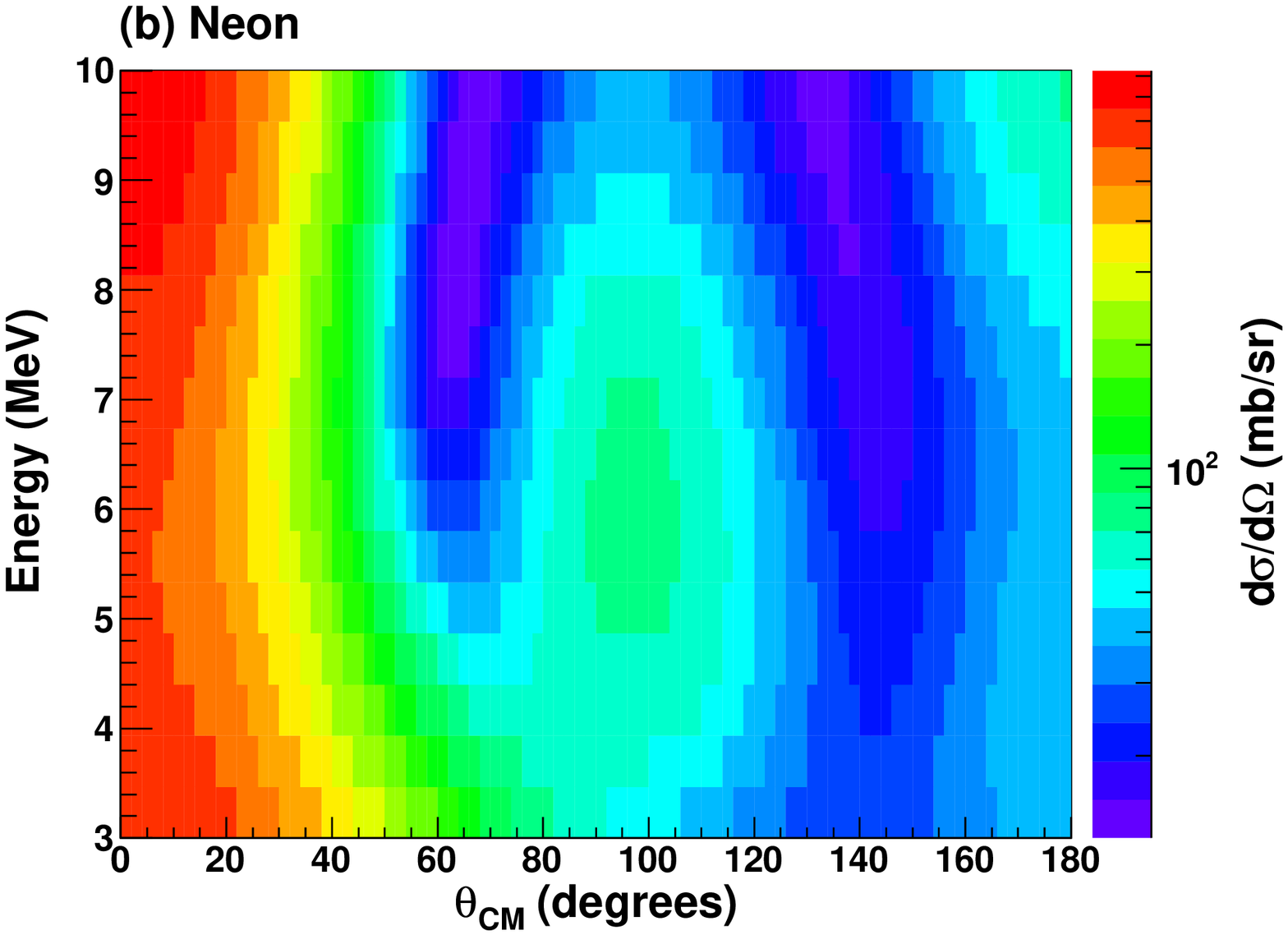}}
  \caption{(Color online) The differential elastic scattering cross section of neutrons from $^{\rm nat}$Ar and $^{\rm nat}$Ne based on an extrapolation of the the optical-model parameter set that best described our data.}               
  \label{fig:cs_2d}
\end{figure}
\end{centering}

The best fit to the cross-section data calculated from the SOM potential (Eqn.~\ref{eqn:opt}) is shown in Figs.~\ref{fig:cs_argon}~--~\ref{fig:cs_2d}. For argon, the total elastic scattering cross section found from the optical-model fit was 2020 mb for $E_n$ = 6.0 MeV and 995 mb for $E_n$ = 14.0 MeV. The cross section was found to differ significantly from ENDF/B-VII.0. This discrepancy is largely due to an over-estimation in the radii ($r_v$,$r_d$,$r_{s.o}$) in the ENDF/B-VII.0 parameterization. Compared to the prediction using the local Koning-Delaroche potential for $^{40}$Ar, the total elastic scattering cross section was found to differ by 8\% at 6.0 MeV and 3\% at 14.0 MeV.

For neon, the total elastic scattering cross section found from the optical-model fits was 1160 mb for $E_n$ = 5.0 MeV and 990 mb for $E_n$ = 8.0 MeV. The cross section was found to differ significantly from the extrapolation of the Dave--Gould potential to the $A=20$ range. The origin of this discrepancy is largely due to the difference in the real volume term ($V_v$) in the optical-model potential. Compared to the prediction using the global Koning-Delaroche potential for $^{20}$Ne, the total elastic scattering cross section was found to differ by 6\% at 5.0 MeV and 13\% at 8.0 MeV. This is the first available data for nuclear masses from \textit{A} = 17 to \textit{A} = 23.

\subsection{Neutron Cross Sections in G\textsc{eant4}}

Many direct dark matter searches rely on Monte-Carlo codes such as G\textsc{eant4}~\cite{Ago03,All06} to simulate experiments and determine background estimates. In G\textsc{eant4}, the \texttt{G4NeutronHP} (high precision) class is used to simulate neutron transport below 20 MeV. It relies on G4NDL, a library of data-driven cross sections and angular distributions, mainly from ENDF-6~\cite{Her09} to simulate neutron processes. In the case where no data exist for a given isotope, \texttt{G4NeutronHP} will replace it with the cross section information from the isotope with the nearest \textit{Z} and \textit{A}~\cite{Men12}. Since cross-section information does not exist for neon in G4NDL library versions up to at least G4NDL 3.14, \texttt{G4NeutronHP} replaces $^{20,22}$Ne with $^{22}$Na, which has completely different nuclear properties. A linear interpolation is done by \texttt{G4NeutronHP} for the data in the G4NDL libraries.

\begin{centering}
\begin{figure}[htp]
  \centering
 \subfloat{\label{fig:argon_sim_6_ang_compare}\includegraphics[width=0.65\textwidth]{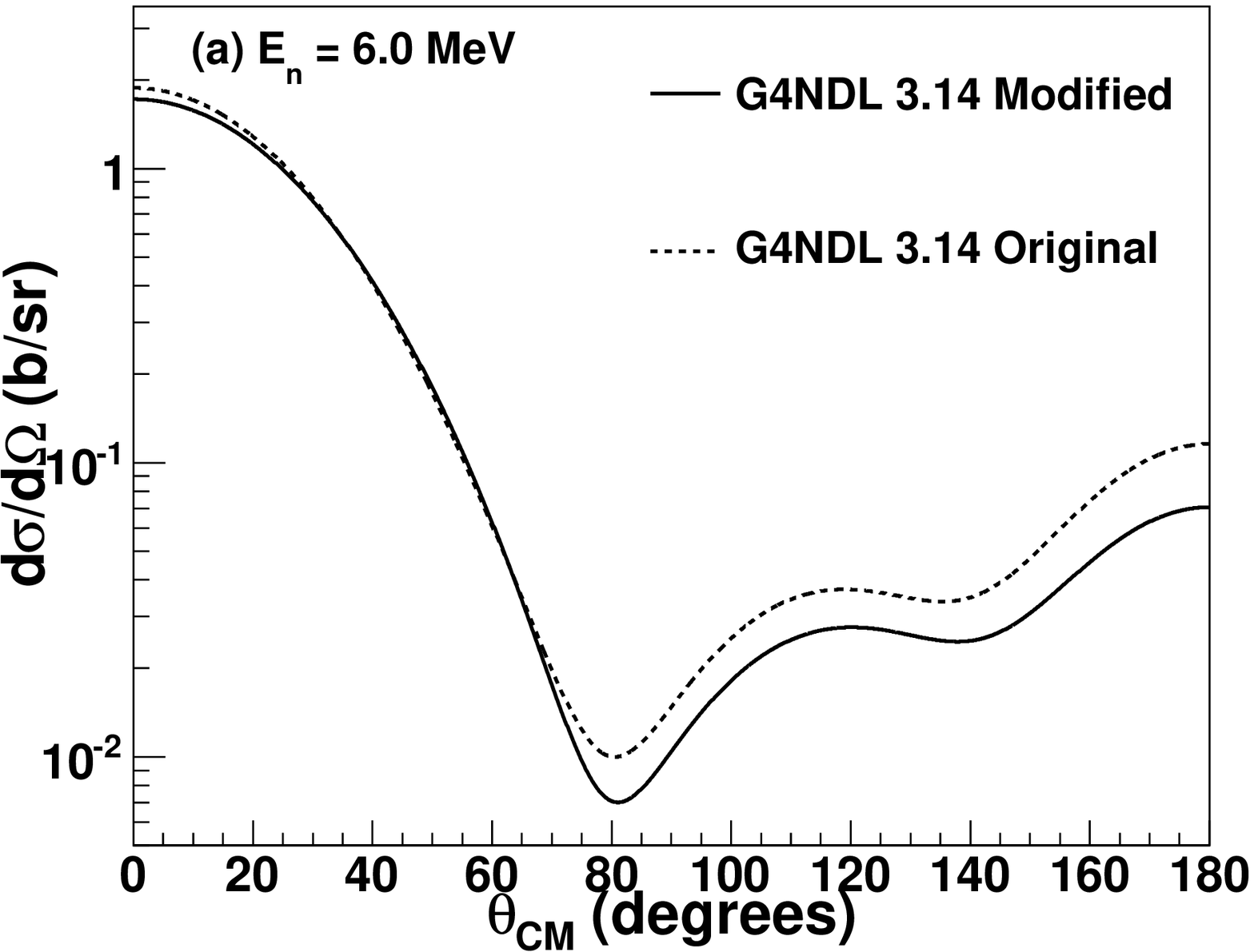}}\\
 \subfloat{\label{fig:argon_sim_14_ang_compare}\includegraphics[width=0.65\textwidth]{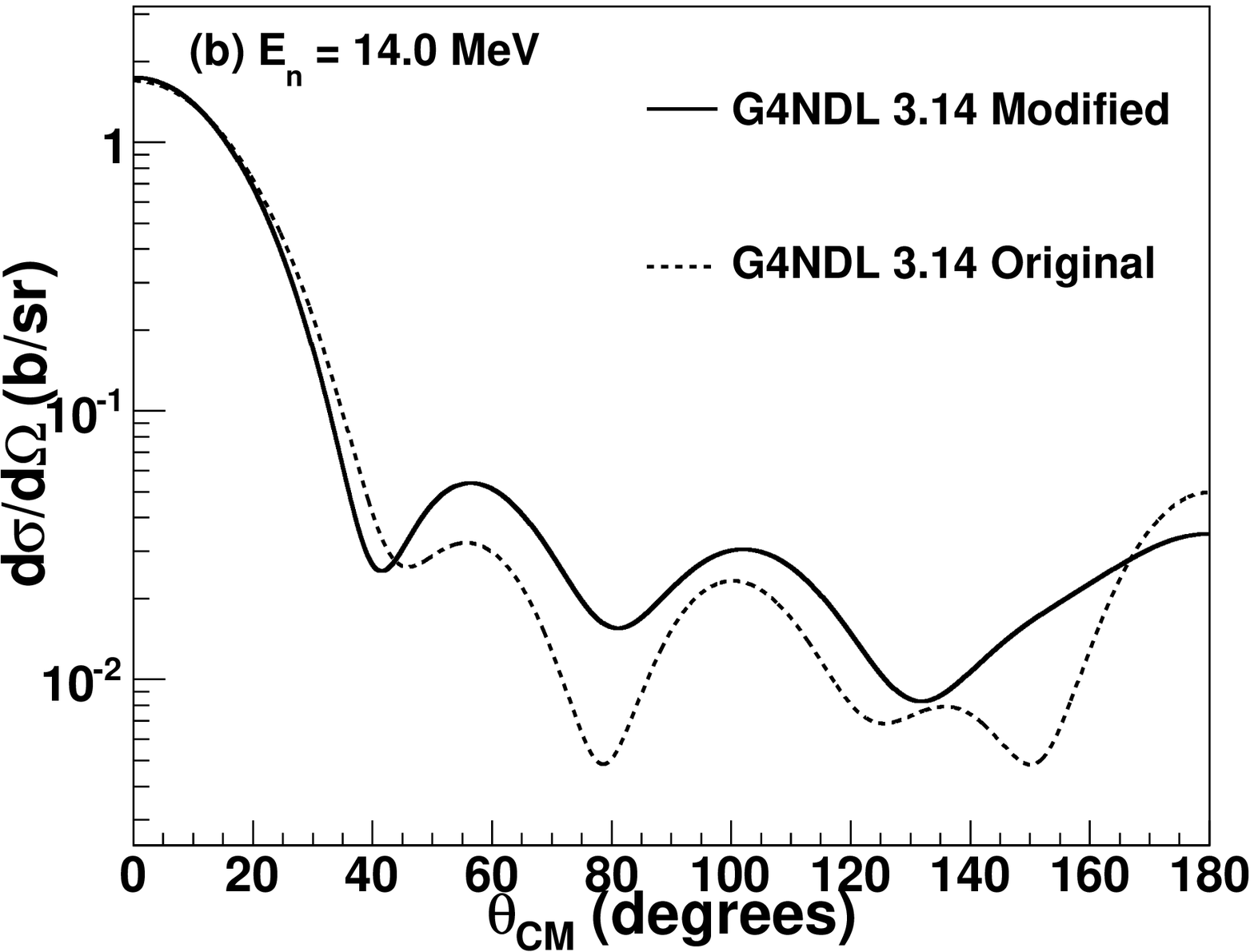}}
  \caption{The solid curves are the differential elastic scattering cross sections of neutrons from $^{40}$Ar at (a) 6.0 and (b) 14.0 MeV added to the G4NDL 3.14 library. The dashed curves are the original G4NDL 3.14 cross sections.}               
  \label{fig:argon_sim_ang_compare}
\end{figure}
\end{centering}

\begin{centering}
\begin{figure}[htp]
  \centering
  \includegraphics[width=0.75\textwidth]{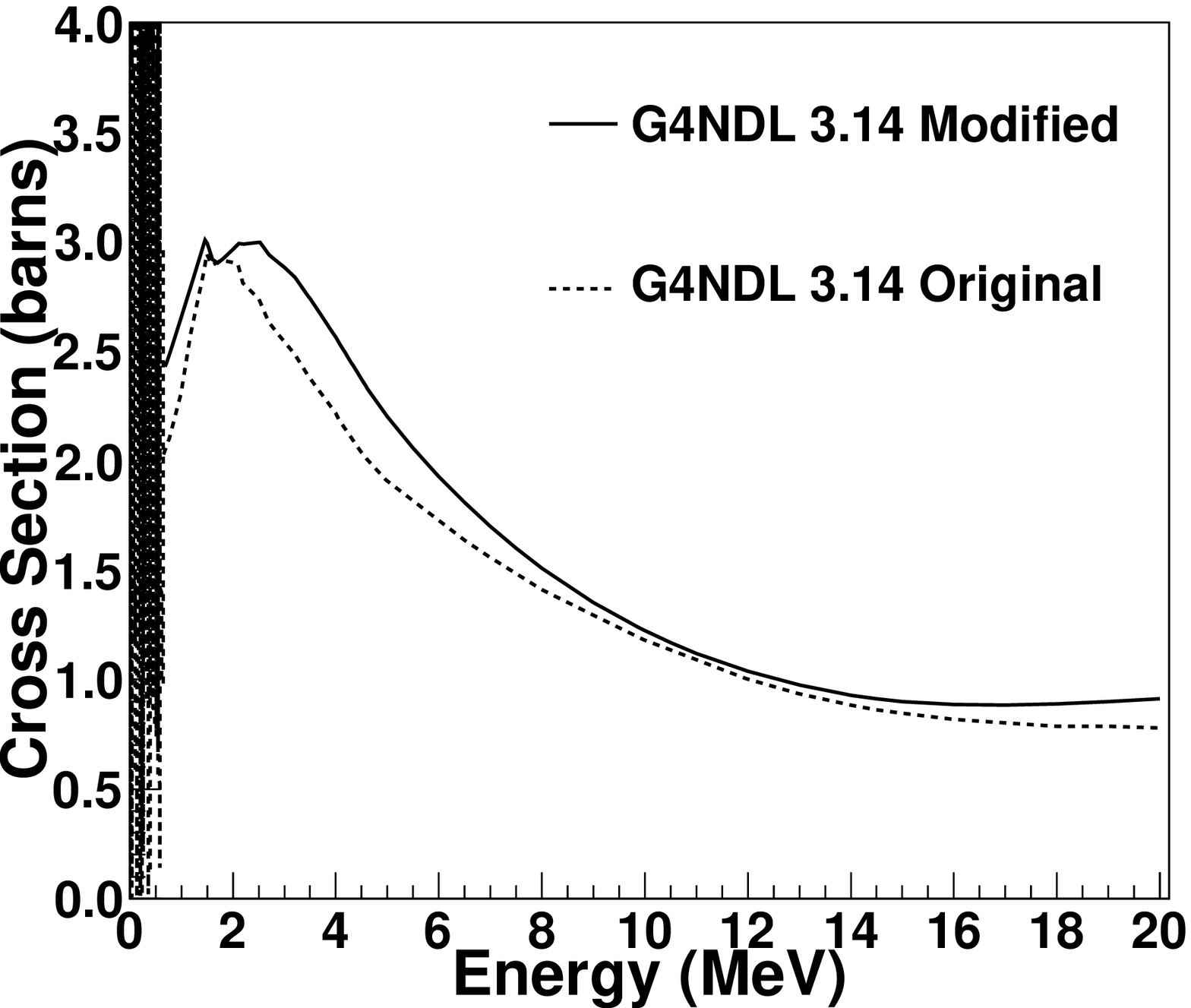}
  \caption{A comparison of the original G4NDL 3.14 elastic scattering cross section and modified cross section for $^{40}$Ar. The cross section shows significant resonance behavior below 650 keV.}               
  \label{fig:argon_sim_tot_compare}
\end{figure}
\end{centering}

We have created data files to add to G4NDL for the cross section and angular distributions for elastic scattering of neutrons up to 20.0 MeV based on our data for argon and neon. For argon, the cross section constructed from our optical-model analysis were used from 0.65 to 20.0 MeV. Below 650 keV, the original cross section, based on the data of Winters {\it et al.}~\cite{Win91}, including the resonances, from the G4NDL 3.14 data file for $^{40}$Ar was used. The original G4NDL 3.14 cross sections and the modified cross sections are shown in Fig.~\ref{fig:argon_sim_ang_compare} for the specific neutron energies analyzed in this work. A comparison of the total elastic scattering cross sections is shown in Fig.~\ref{fig:argon_sim_tot_compare}.

\begin{centering}
\begin{figure}[htp]
  \centering
  \subfloat{\label{fig:neon_sim_5_ang_compare}\includegraphics[width=0.65\textwidth]{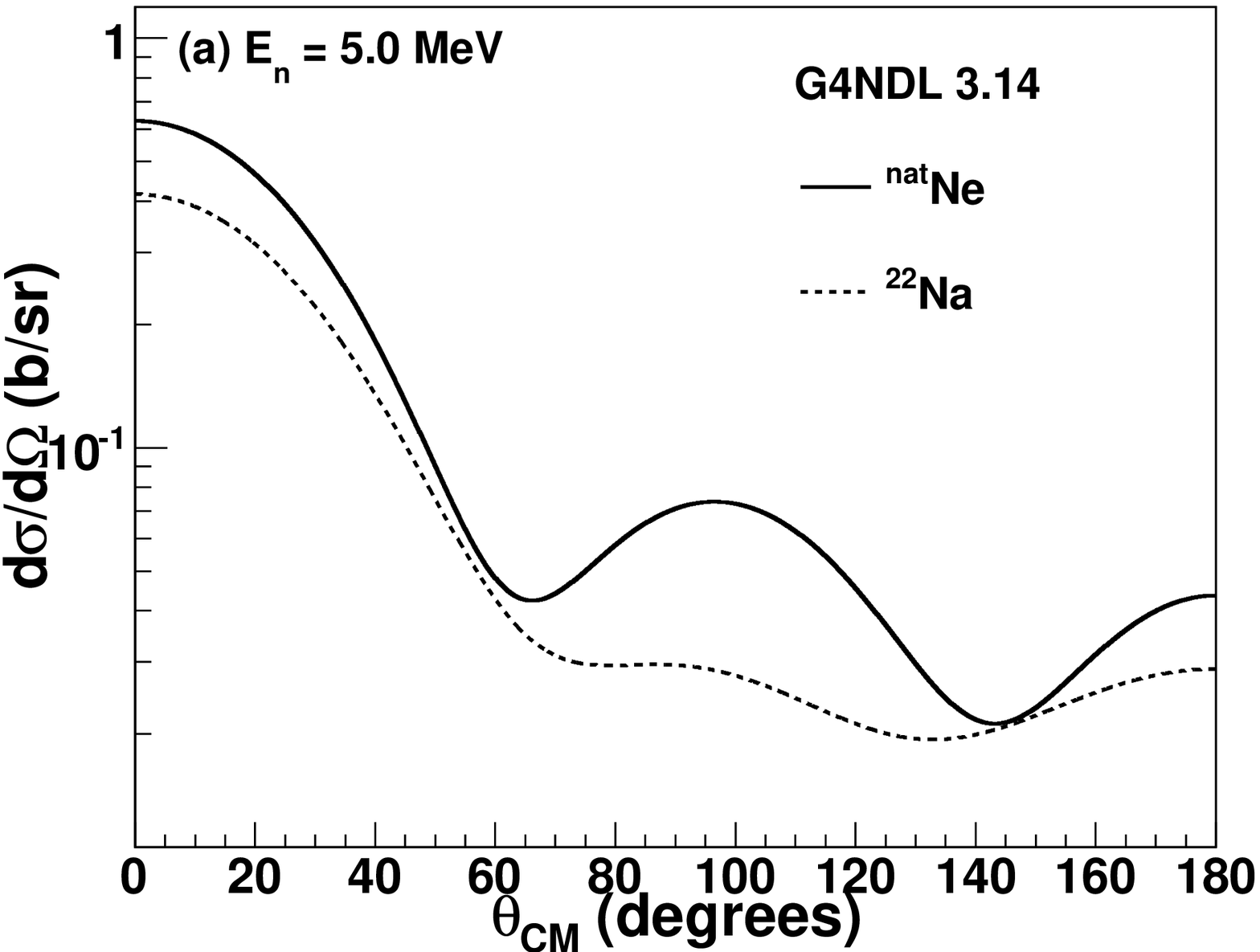}} \\
  \subfloat{\label{fig:neon_sim_8_ang_compare}\includegraphics[width=0.65\textwidth]{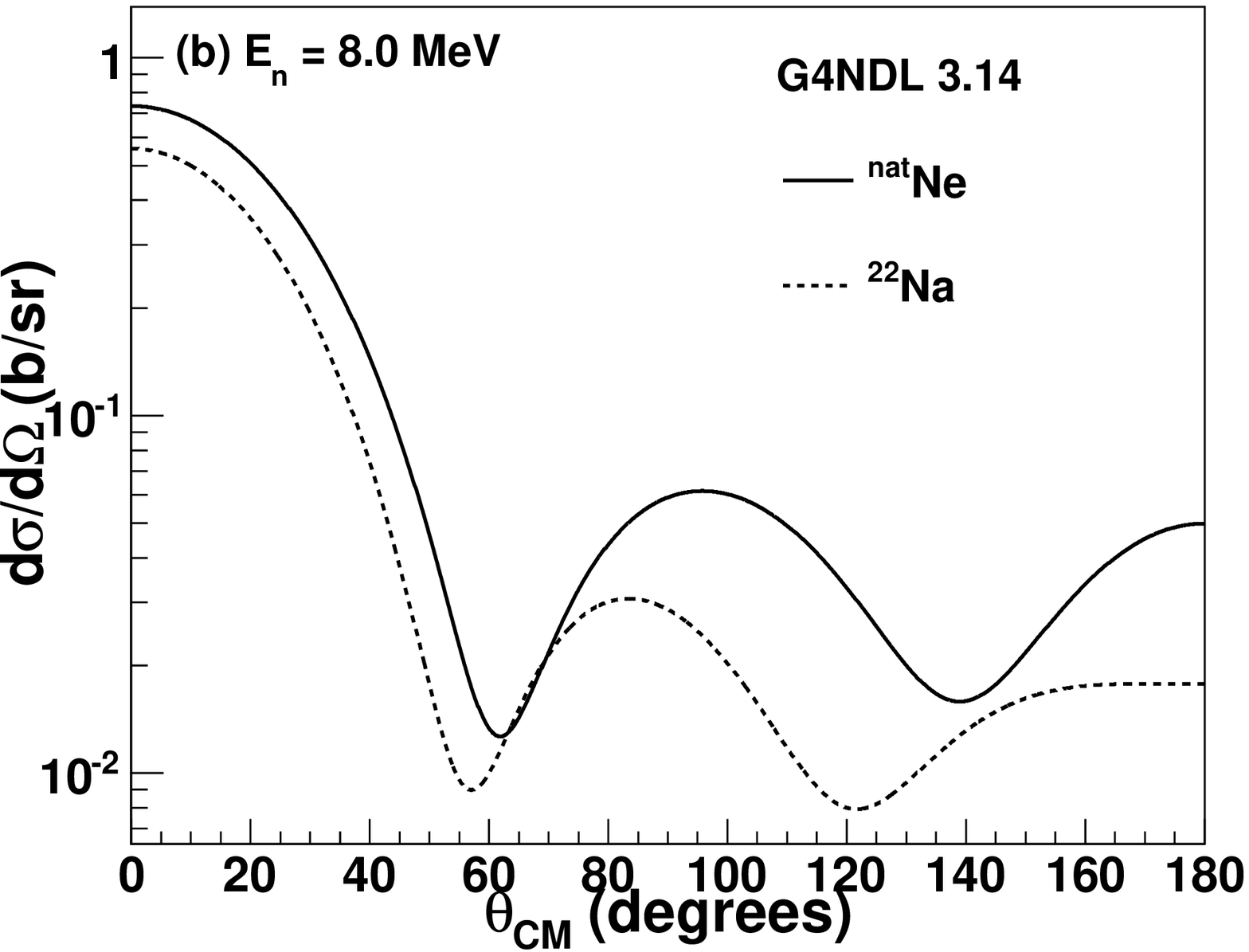}}
  \caption{The solid curves are the differential elastic scattering cross sections of neutrons from $^{\rm nat}$Ne at (a) 5.0 and (b) 8.0 MeV added to the G4NDL 3.14 library. The dashed curves are the original G4NDL 3.14 cross sections for $^{22}$Na.}               
  \label{fig:neon_sim_ang_compare}
\end{figure}
\end{centering}

\begin{centering}
\begin{figure}[htp]
  \centering
  \includegraphics[width=0.75\textwidth]{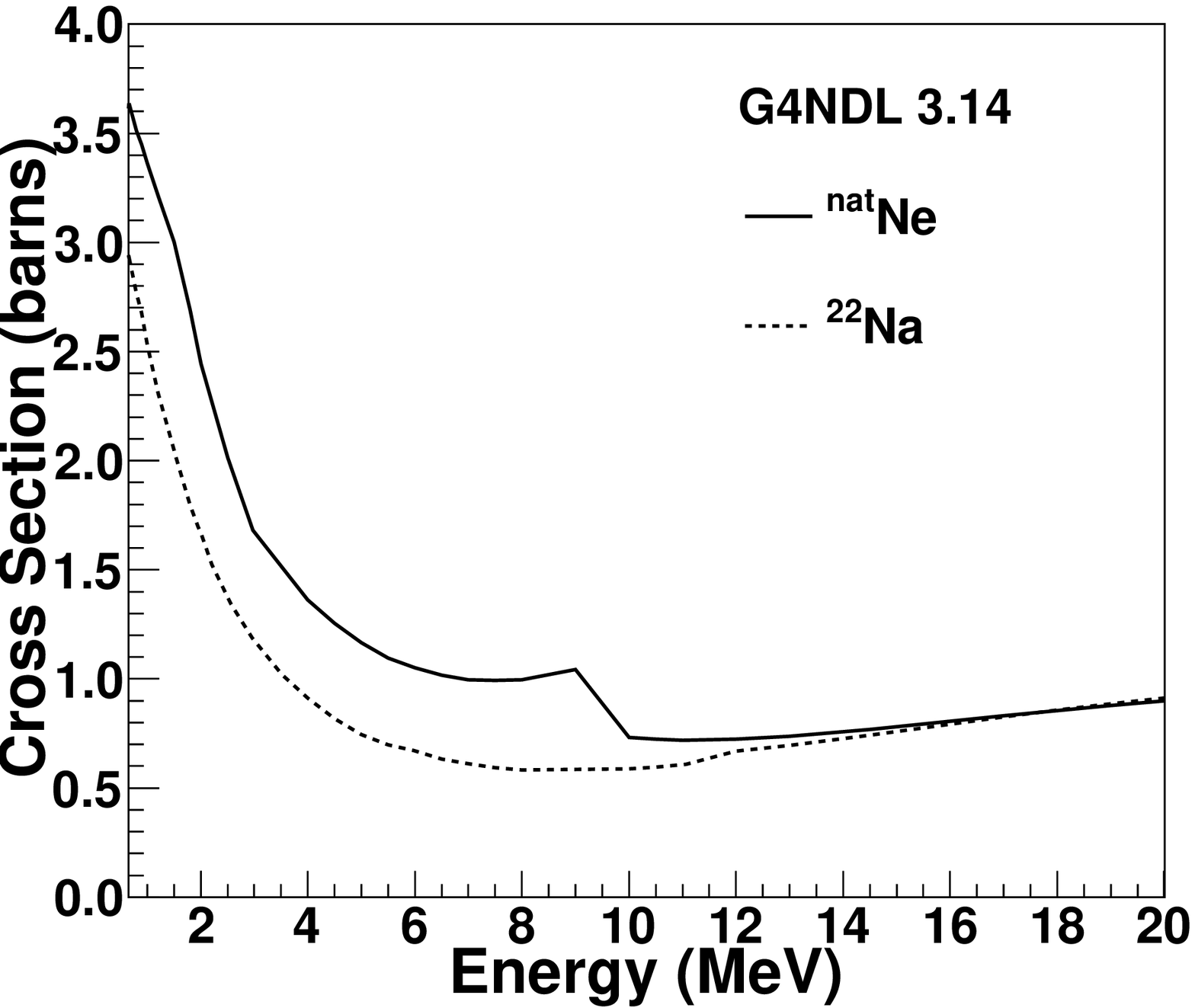}
  \caption{A comparison of the original G4NDL 3.14 elastic scattering cross section for $^{22}$Na and modified cross section for $^{\rm nat}$Ne. The solid curve is the elastic scattering cross section of neutrons from $^{\rm nat}$Ne added to the G4NDL 3.14 library. The dashed curve is the original G4NDL 3.14 cross section for $^{22}$Na.}               
  \label{fig:neon_sim_tot_compare}
\end{figure}
\end{centering}

For neon, for reasons discussed above, we chose to restrict our optical-model analysis to between 3.0 and 10.0 MeV. Below 3.0 MeV and from 10.0 to 20.0 MeV, the cross section calculated from the Koning--Delaroche global potential~\cite{Kon03} was used. The original G4NDL 3.14 cross section and the modified cross section is shown in Fig.~\ref{fig:neon_sim_ang_compare} for the specific neutron energies analyzed in this work. A comparison of the total elastic scattering cross section is shown in Fig.~\ref{fig:neon_sim_tot_compare}. At 3.0 MeV, the transition from the cross section using the global potential and the fits from this work is smooth. There is a kink in the cross section at 10.0 MeV, where the cross section transitions back to the calculation from the global potential. We conclude that although the cross-section calculation from the Koning--Delaroche global potential is better than the current $^{22}$Na data file used by \texttt{G4NeutronHP}, is not adequate to describe the cross section between 10.0 and 20.0 MeV. Additional measurements would be necessary to constrain our model at energies above 10.0 MeV based on our data. The cross section in this energy range do not affect dark matter background estimates from $^{238}$U- and $^{232}$Th-induced ($\alpha$,n) reactions because the neutron spectrum cuts off sharply at about 8 MeV~\cite{Mei09}.

\section{Conclusions}

The differential elastic scattering cross section was measured for 5.0- and 8.0-MeV neutrons incident on neon and for 6.0-MeV neutrons incident on argon. These data, along with the 14.0-MeV data for argon from Beach~{\it et al.}~\cite{Bea67} were fit using the spherical optical model and compared to global optical-model predictions. Precise neutron scattering cross sections are required to estimate neutron backgrounds from ($\alpha$,n) reactions in direct dark matter searches. These cross sections, which were previously unmeasured, are an important component in background estimates of these experiments because the simulation relies on available data in the energy range below 20 MeV. Our data significantly improve the current cross-section libraries in the \textsc{Geant4} simulation code, add to the nuclear databases, and provide accurate data for benchmarking global optical models.

\section{Acknowledgements}
This work was primarily supported by DOE grants \# DE-FG02-97ER4104 and \# DE-FG02-97ER41033. We thank R. Reifarth for providing us with the empty gas cells and the associated filling adapters. We also acknowledge the support of the TUNL technical staff.


\appendix*

\section{Elastic scattering cross-section data for neutrons from argon and neon}

\begin{table*}[htp]
    \centering
    \caption{Elastic scattering cross-section data for 5.0-MeV neutrons from $^{\rm nat}$Ne.}
    \begin{tabular}{l l l l l} 
    \hline
    \hline
    \multicolumn{3}{l}{E$_{n}$ = 5.0 $\pm$ 0.4 MeV} \\
    \multicolumn{3}{l}{$\sigma_{el}$ = 1290 $\pm$ 40 mb} \\
    \multicolumn{3}{l}{d$\sigma(0^{\degree})$/d$\Omega$ = 550 $\pm$ 30 mb/sr} \\
    \hline
    $\theta_{c.m.}$ & (d$\sigma$/d$\Omega$)$_{data}$ & $\Delta$(d$\sigma$/d$\Omega$)$_{data}$ & (d$\sigma$/d$\Omega$)$_{fit}$ & $\chi_{\rm fit}^2/{\textrm{data point}}$ \\
    (degrees) & (mb/sr) & (mb/sr) & (mb/sr) &  \\
    \hline
    0          &     535 & 50       &          549      &          0.08     \\
    23          &     373 & 76       &          409      &          0.2     \\
    31          &     308 & 37       &          320      &          0.09     \\
    39          &     228 & 21       &          232      &          0.03     \\
    47          &     201 & 18       &          157      &          5.7     \\
    52          &     109 & 18       &          120      &          0.3     \\
    63          &     71 & 5       &          71      &          0.005     \\
    73          &     56 & 4       &          60      &          0.9     \\
    83          &     80 & 6       &          67      &          3.7     \\
    93          &     72 & 6       &          75      &          0.2     \\
    103          &     83 & 7       &          75      &          1.3     \\
    113          &     71 & 5       &          63      &          2.2     \\
    122          &     38 & 3       &          47      &          6.0     \\
    134          &     30 & 4       &          27      &          0.7     \\
    142          &     27 & 3       &          20      &          4.4     \\
    149          &     16 & 4       &          19      &          0.5     \\
    156          &     20 & 4       &          23      &          0.2     \\
    \hline
    \hline
    \end{tabular}
    \label{tab:ntof_neon5_data}
\end{table*}

\begin{table*}[htp]
    \centering
    \caption{Elastic scattering cross-section data for 8.0-MeV neutrons from $^{\rm nat}$Ne.}
    \begin{tabular}{l l l l l} 
    \hline
    \hline
    \multicolumn{3}{l}{E$_{n}$ = 8.0 $\pm$ 0.4 MeV} \\
    \multicolumn{3}{l}{$\sigma_{el}$ = 940 $\pm$ 30 mb} \\
    \multicolumn{3}{l}{d$\sigma(0^{\degree})$/d$\Omega$ = 710 $\pm$ 40 mb/sr} \\
    \hline
    $\theta_{c.m.}$ & (d$\sigma$/d$\Omega$)$_{data}$ & $\Delta$(d$\sigma$/d$\Omega$)$_{data}$ & (d$\sigma$/d$\Omega$)$_{fit}$ & $\chi_{\rm fit}^2/{\textrm{data point}}$ \\
    (degrees) & (mb/sr) & (mb/sr) & (mb/sr) &  \\
    \hline
    0          &     670 & 70       &          706      &          0.3     \\
    29          &     329 & 33       &          315      &          0.2     \\
    34          &     233 & 18       &          225      &          0.2     \\
    38          &     153 & 13       &          162      &          0.4     \\
    42          &     122 & 10       &          110      &          1.3     \\
    46          &     61 & 6       &          68      &          1.4     \\
    50          &     41 & 4       &          39      &          0.3     \\
    54          &     27 & 4       &          20      &          2.7     \\
    60          &     7 & 2       &          8      &          0.1     \\
    65          &     12 & 2       &          10      &          0.9     \\
    69          &     14 & 1       &          17      &          3.2     \\
    73          &     35 & 2       &          25      &          12.1     \\
    83          &     43 & 4       &          45      &          0.2     \\
    93          &     58 & 5       &          55      &          0.4     \\
    103          &     49 & 3       &          53      &          0.9     \\
    113          &     44 & 3       &          41      &          1.2     \\
    123          &     25 & 1       &          25      &          0.01     \\
    134          &     14 & 1       &          13      &          1.0     \\
    145          &     13 & 1       &          14      &          0.04     \\
    155          &     29 & 2       &          28      &          0.4     \\
    \hline
    \hline
    \end{tabular}
    \label{tab:ntof_neon8_data}
\end{table*}

\begin{table*}[htp]
    \centering
    \caption{Elastic scattering cross-section data for 6.0 MeV neutrons from $^{\rm nat}$Ar.}
    \begin{tabular}{l l l l l} 
    \hline
    \hline
    \multicolumn{3}{l}{$E_{n}$ = 6.0 $\pm$ 0.4 MeV} \\
    \multicolumn{3}{l}{$\sigma_{el}$ = 2170 $\pm$ 60 mb} \\
    \multicolumn{3}{l}{$\sigma(0^{\circ})$= 1840 $\pm$ 130 mb/sr} \\
    \hline
    $\theta_{CM}$ & $\sigma(\theta)_{\rm data}$ & $\Delta \sigma(\theta)_{\rm data}$ & $\sigma(\theta)_{\rm fit}$ & $\chi_{\rm fit}^2/{\rm data point}$ \\
    (degrees) & (mb/sr) & (mb/sr) & (mb/sr) &  \\
    \hline
    0          &     1650 & 165       &          1842      &          1.4     \\
    23          &     1354 & 75       &          1268      &          1.3     \\
    33          &     893 & 70       &          796      &          1.9     \\
    38          &     567 & 33       &          575      &          0.05     \\
    43          &     380 & 23       &          386      &          0.05     \\
    48          &     218 & 14       &          241      &          2.5     \\
    53          &     148 & 10       &          139      &          0.8     \\
    61          &     56 & 4       &          51      &          1.7     \\
    71          &     13 & 1       &          14      &          0.04     \\
    81          &     6 & 1       &          6      &          0.5     \\
    91          &     9 & 1       &          9      &          0.08     \\
    101          &     20 & 1       &          20      &          0.003    \\
    111          &     28 & 1       &          27      &          0.6     \\
    122          &     25 & 2       &          27      &          0.2     \\
    133          &     28 & 2       &          27      &          0.3     \\
    144          &     30 & 2       &          30      &          0.001     \\
    155          &     39 & 3       &          39      &          0.02     \\
    \hline
    \hline
    \end{tabular}
    \label{tab:ntof_argon6_data}
\end{table*}

\begin{table*}[htp]
    \centering
    \caption{Elastic scattering cross-section data for 14.0 MeV neutrons from $^{\rm nat}$Ar. These data were taken from Ref.\cite{Bea67} and fit using the same procedure as the TUNL data.}
    \begin{tabular}{l l l l l} 
    \hline
    \hline
    \multicolumn{3}{l}{$E_{n}$ = 14.0 $\pm$ 0.4 MeV} \\
    \multicolumn{3}{l}{$\sigma_{el}$ = 975 $\pm$ 20 mb} \\
    \multicolumn{3}{l}{$\sigma(0^{\circ})$= 1600 $\pm$ 30 mb/sr} \\
    \hline
    $\theta_{CM}$ & $\sigma(\theta)_{\rm data}$ & $\Delta \sigma(\theta)_{\rm data}$ & $\sigma(\theta)_{\rm fit}$ & $\chi_{\rm fit}^2/{\rm data point}$ \\
    (degrees) & (mb/sr) & (mb/sr) & (mb/sr) &  \\
    \hline
    0         &     1880 & 190         &      1604            &        2.1     \\
    10         &      1378 & 66     &          1290           &         1.8     \\
    15         &      987 & 72      &         976             &       0.02     \\
    20      &         645 & 24      &         648             &       0.02     \\
    26      &         353 & 14      &         367             &       1.07     \\
    31      &         180 & 7       &        179              &      0.02     \\
    36      &         80 & 4        &       79                &    0.08     \\
    41      &        46 & 3         &      47                 &   0.06     \\
    46      &         69 & 3        &       51                &    29.0     \\
    51      &         51 & 3        &       63                &    24.0     \\
    56      &         77 & 4        &       68                &    6.0     \\
    61      &         61 & 3        &       62                &    0.1     \\
    66      &         61 & 3        &       48                &    14.5     \\
    71      &         34 & 2        &      34                 &   0.05     \\
    76      &         21 & 2        &       23                &    1.09     \\
    81      &         19 & 2        &       17                &    1.2     \\
    91      &         22 & 1        &       21                &    0.1     \\
    101     &          30 & 3       &        29               &    0.1     \\
    111     &          24 & 2       &        24               &     0.04     \\
    121     &          12 & 1       &        12               &     0.001     \\
    131     &          9.4 & 1.1        &       8                 &   1.6     \\
    141     &          11 & 1       &        11               &     0.03     \\
    151     &          12 & 1       &        11               &     0.6     \\
    159     &          13 & 2       &        13               &     0.05     \\
    \hline
    \hline
    \end{tabular}
    \label{tab:ntof_argon14_data}
\end{table*}

\end{document}